\documentclass[pra,aps,twocolumn,nopacs,superscriptaddress,nofootinbib,longbibliography]{revtex4-1} 
\usepackage{amsmath}  \usepackage{amssymb}  \usepackage{amsfonts}  \usepackage{bm}  \usepackage{bbm}   \usepackage{bbold}  \usepackage{braket}  \usepackage{comment}  \usepackage{dcolumn}  \usepackage{enumerate}  \usepackage{gensymb}  \usepackage{graphicx}  \usepackage{indentfirst}  \usepackage{lmodern}  \usepackage{mathrsfs}  \usepackage{mathtools}  \usepackage{soul}  \usepackage{xcolor}  \usepackage{float}  \usepackage[colorlinks=true,linkcolor=blue,citecolor=blue,urlcolor=blue]{hyperref}  \usepackage[T1]{fontenc}

\def\ii{{\rm i}}  \def\ee{{\rm e}}

        \def\Eb{{\bf E}}                                      \def\rb{{\bf r}}       
                
\def\EF{{E_{\rm F}}}     
        
\def\ii{{i}}  \def\ee{{e}}
\def\um{{\mathrm{\text\textmu m}}}
\def\us{{\mathrm{\text\textmu s}}}

\begin{document}
\title{Plasmon-Enhanced Second-Harmonic Generation in Atomically Thin Crystalline Silver Nanostructures}

\author{Saad~Abdullah} 
\thanks{These authors contributed equally to this work.}
\affiliation{ICFO--Institut de Ciencies Fotoniques, The Barcelona Institute of Science and Technology, 08860 Castelldefels (Barcelona), Spain}

\author{Philipp~K.~Jenke} 
\thanks{These authors contributed equally to this work.}
\affiliation{University of Vienna, Faculty of Physics, Vienna Center for Quantum Science and Technology (VCQ), Boltzmanngasse 5, 1090 Vienna, Austria}

\author{Andrew~P. Weber} 
\affiliation{ICFO--Institut de Ciencies Fotoniques, The Barcelona Institute of Science and Technology, 08860 Castelldefels (Barcelona), Spain}

\author{\'{A}lvaro~Rodr\'{i}guez~Echarri} 
\altaffiliation[Present address: ]{Center for Nanophotonics, NWO Institute AMOLF, 1098 XG Amsterdam, The Netherlands}
\affiliation{ICFO--Institut de Ciencies Fotoniques, The Barcelona Institute of Science and Technology, 08860 Castelldefels (Barcelona), Spain}

\author{Vahagn~Mkhitaryan} 
\affiliation{ICFO--Institut de Ciencies Fotoniques, The Barcelona Institute of Science and Technology, 08860 Castelldefels (Barcelona), Spain}

\author{Fadil~Iyikanat} 
\altaffiliation[Present address: ]{Department of Physics, Dokuz Eyl\"ul University, 35390 Izmir, Turkey}
\affiliation{ICFO--Institut de Ciencies Fotoniques, The Barcelona Institute of Science and Technology, 08860 Castelldefels (Barcelona), Spain}

\author{Frederik~Schiller} 
\affiliation{Centro de F\'{\i}sica de Materiales CSIC-UPV/EHU and Materials Physics Center, 20018 San Sebastian, Spain}

\author{Philip~Walther} 
\affiliation{University of Vienna, Faculty of Physics, Vienna Center for Quantum Science and Technology (VCQ), Boltzmanngasse 5, 1090 Vienna, Austria}
\affiliation{University of Vienna, Research Platform for Testing the Quantum and Gravity Interface (TURIS), Boltzmanngasse 5, 1090 Vienna, Austria, and Christian Doppler Laboratory for Photonic Quantum Computer, Faculty of Physics, University of Vienna, 1090 Vienna, Austria}

\author{J.~Enrique~Ortega} 
\affiliation{Centro de F\'{\i}sica de Materiales CSIC-UPV/EHU and Materials Physics Center, 20018 San Sebastian, Spain}
\affiliation{Departamento de F\'{\i}sica Aplicada, Universidad del Pa\'{\i}s Vasco/EHU, 20018 San Sebastian, Spain}

\author{Lee~A.~Rozema} 
\email[Corresponding author: ]{lee.rozema@univie.ac.at}
\affiliation{University of Vienna, Faculty of Physics, Vienna Center for Quantum Science and Technology (VCQ), Boltzmanngasse 5, 1090 Vienna, Austria}

\author{F.~Javier Garc\'{\i}a~de~Abajo} 
\email[Corresponding author: ]{javier.garciadeabajo@nanophotonics.es}
\affiliation{ICFO--Institut de Ciencies Fotoniques, The Barcelona Institute of Science and Technology, 08860 Castelldefels (Barcelona), Spain}
\affiliation{ICREA--Instituci\'o Catalana de Recerca i Estudis Avan\c{c}ats, Passeig Llu\'{\i}s Companys 23, 08010 Barcelona, Spain}

\begin{abstract}
The intrinsically weak nonlinear optical response of existing materials, further constrained by symmetry-forbidden second-order processes in centrosymmetric media, severely limits efficient frequency conversion in deeply subwavelength, ultrathin volumes. Addressing this challenge is crucial for the development of nonlinear nanophotonics. Here, we show that atomically thin, epitaxially grown crystalline silver films circumvent these restrictions through the interplay of vertical electronic quantum confinement and lateral plasmonic enhancement. We fabricate atomically thin films that exhibit an enhanced nonlinear response associated with electronic quantum wells, and subsequently pattern them into periodic nanoribbon and nanotriangle arrays sustaining infrared localized surface plasmon resonances. Strong near-field confinement in these structures further boosts second-harmonic generation compared to unpatterned films. Precise control over nanostructure geometry enables spectral tuning of the plasmonic resonance, and consequently, the enhanced harmonic frequency. Our findings establish an approach for activating robust second-order nonlinearities in quantum-confined metals, where intrinsic size effects and plasmonic resonances act synergistically. The compatibility of high-quality epitaxial growth with microchip fabrication technology offers a scalable route toward ultracompact nonlinear optical components for on-chip frequency conversion, sensing, and quantum photonic applications.
\end{abstract}
\maketitle

\section{Introduction} \label{section:Introduction}

Since the discovery of nonlinear optical phenomena, considerable effort has been devoted to identifying materials that display substantial nonlinear responses at low optical intensities \cite{SMZ19,MSA22,AK25}. Traditionally, second-harmonic generation (SHG) has been realized using bulk non-centrosymmetric crystals such as lithium niobate \cite{LHL18,QWC25}, potassium titanyl phosphate \cite{YYY95}, or beta-barium borate \cite{B08_3}, where symmetry breaking in their crystal structures enables efficient second-order nonlinearities. While these materials exhibit comparatively strong nonlinear coefficients, their large interaction volumes and stringent fabrication constraints (including phase-matching requirements) limit their seamless integration with on-chip photonic nanodevices.

To address this challenge, a broad range of alternative SHG platforms has been explored, including plasmonic metamaterials \cite{KZ12,LSJ14,WJW21}, two-dimensional (2D) materials \cite{LLG23,KFG19,TFY25}, quantum dots \cite{ZOC09,RCL21}, and other emerging nonlinear nanophotonic systems \cite{WRA09,BSD14,DZW17}. Efforts along these directions have focused on the development of non-centrosymmetric materials and resonant nanophotonic architectures capable of enhancing nonlinear light--matter interactions.

Recent years have witnessed major advances in ultrathin nonlinear optical platforms, including layered ferroelectrics, rhombohedral van der Waals crystals, dielectric metasurfaces, and hybrid resonant nanostructures. For example, ferroelectric NbOI$_2$ nanosheets have demonstrated effective nonlinear susceptibilities approaching the $10^3$~pm/V range together with peak-intensity-normalized SHG conversion efficiencies exceeding 0.2\% under resonant excitation conditions \cite{ATW22}, while rhombohedral 3R-MoS$_2$ structures have exhibited resonantly enhanced nonlinear susceptibilities approaching $\sim800$~pm/V together with SHG enhancement factors exceeding two orders of magnitude in resonant nanodisk geometries \cite{ZAY24}. More recently, periodically poled layered semiconductor structures based on 3R-MoS$_2$ have enabled quasi-phase-matched nonlinear frequency conversion efficiencies approaching the 0.1\% range over micrometre-scale interaction lengths \cite{TFY25}. Likewise, dielectric metasurfaces and quasi-bound-state-in-the-continuum architectures integrated with layered semiconductors have demonstrated SHG enhancement factors exceeding $10^3$ through strong resonant field confinement \cite{BKW20}.

While each of these novel nonlinear platforms offers distinct advantages and enhancement mechanisms, atomically thin crystalline noble metals provide access to a complementary nonlinear nanophotonic regime in which conduction electrons are vertically confined, giving rise to quantum finite-size effects and the formation of quantum-well states (QWS) that modify the metallic electronic band structure and can enhance the nonlinear optical response \cite{PPK99,HKT01,PPM06,paper382,PTQ24,paper465}. In parallel, plasmonic resonances supported by such ultrathin crystalline metallic films sustain high-quality, deep-subwavelength optical confinement at visible and near-infrared frequencies \cite{paper335,paper427}. Combined with additional lateral field enhancement enabled through nanopatterning, these ultrathin crystalline metallic nanostructures support enhanced SHG together with strong optical near-field confinement, tunable plasmonic resonances, scalable fabrication, and direct compatibility with silicon-based nanophotonic architectures.

The geometry and size of nanostructures critically influence their plasmonic behavior \cite{S11}. Precise control over these parameters is essential to ensure that localized plasmon resonances match the excitation conditions required for efficient nonlinear conversion. A variety of techniques have been developed to meet such fabrication demands, including ion-beam milling \cite{CPB10}, colloidal synthesis \cite{LCM13}, self-assembly organization \cite{AKT17}, electrochemical deposition \cite{CWZ19}, and vapor deposition methods \cite{PTT15}, among others. While each approach offers distinct advantages, they often fall short in resolution, spatial control, or substrate compatibility when applied to subnanometer-scale volumes. In addition, achieving high-quality, scalable, and tunable plasmonic resonances that enhance the local field strength also requires minimizing losses, particularly in metallic films, which often require tuning the deposition parameters \cite{MJK15}.

\begin{figure*}
\centering \includegraphics[width=1.0\textwidth]{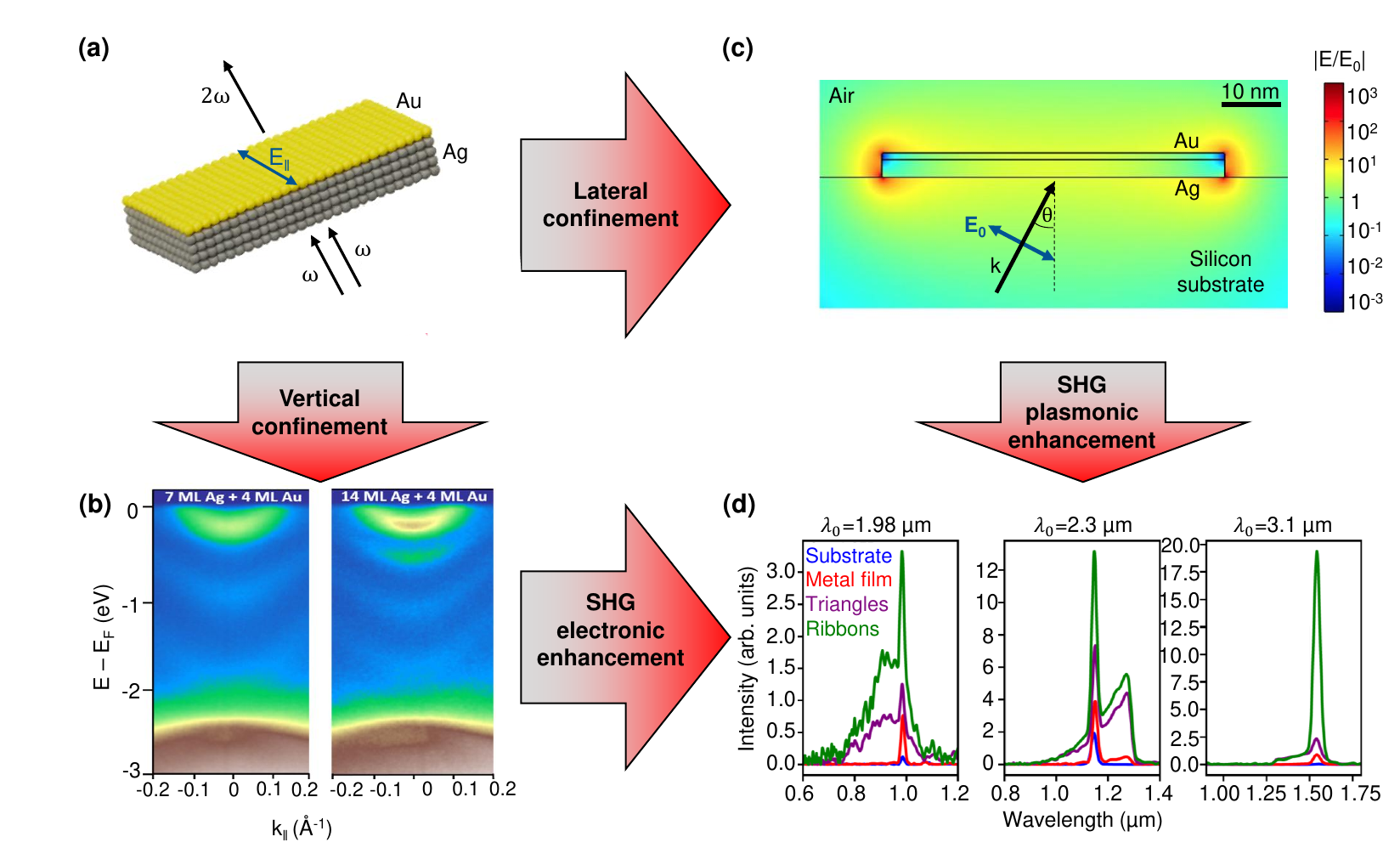}
\caption{{\bf Plasmon-enhanced second-harmonic generation (SHG) in ultrathin crystalline silver films.} General scheme illustrating the generation of an enhanced second-harmonic signal driven by vertical electronic confinement and lateral plasmonic near-field enhancement. 
\textbf{(a)}~We consider atomically thin crystalline metal nanoribbons capped with gold. Pairs of incident photons at the fundamental frequency $\omega$ are converted into second-harmonic photons at frequency $2\omega$. The input and output light polarizations are oriented across the ribbons.
\textbf{(b)}~Dispersion diagrams of electronic quantum wells in the metal revealed through angle-resolved photoemission spectroscopy (ARPES) performed on films consisting of 7 and 14 Ag(111) monolayers (MLs), capped with 4~MLs of Au(111) each.
\textbf{(c)}~Electric near-field enhancement in response to plane-wave illumination calculated for p-polarized light incident with an angle $\theta\approx13^\circ$ and field amplitude $\Eb_0$.
\textbf{(d)}~Measured SHG intensity spectra at different excitation wavelengths $\lambda_0$ (see labels) for various nanostructure geometries, including the bare silicon substrate and the unstructured metal film for reference, as well as periodic ribbon and triangle arrays. The metal thickness is 7+4 MLs of Ag+Au. The ribbon width and triangle side are $W=150~$nm for $\lambda_0=3.1~\um$ and $W=110~$nm for the two other wavelengths, while the period is $3\,W$ in both cases.}
\label{Fig1}
\end{figure*}

In this work, we overcome fabrication challenges by combining epitaxially grown ultrathin silver films with precise lateral patterning of periodic nanostructures \cite{paper335,paper427} to achieve an efficient nonlinear optical response. Using high-resolution electron-beam lithography (EBL), we fabricate well-defined planar arrays of crystalline silver nanostructures that enable tunable plasmonic resonances featuring high quality factors and deep-subwavelength confinement. Large boosts in SHG are observed, driven by vertical electronic confinement and lateral plasmonic enhancement associated with their atomic-scale thickness (Figure~\ref{Fig1}). Specifically, we fabricate crystalline silver ribbons and triangles with thicknesses ranging from 7 to 14 monolayers (MLs) (i.e., $\sim 1.7-3.3$~nm thickness), (111) surface orientation, widths in the $W=50-400$~nm range, and periods $P=3\,W$. The epitaxial growth of the metal film, followed by surface passivation with $3-4$ MLs of gold, bears critical importance to minimize the density of defects and grain boundaries, resulting in high plasmon quality, low optical damping, and enhanced SHG. Our results establish a promising strategy for integrating strong nonlinearities into nanophotonic architectures, advancing the long-standing goal of realizing efficient nonlinear optical devices.

\section{Results} \label{section:Discussion}

We focus on the nonlinear optical properties of atomically thin (few-monolayer) epitaxially grown crystalline silver films of (111) surface orientation capped with a few monolayers of gold (see Methods). The gold capping layer, which closely matches the lattice constant of silver, grows epitaxially and enhances the structural stability of the heterostructure while preventing oxidation of the underlying silver. The growth protocol is optimized to produce high-quality crystalline films, as explained elsewhere \cite{paper335}. Achieving ultrathin noble-metal layers that are simultaneously stable and possess high crystalline quality is generally not feasible with conventional thin-film deposition techniques such as sputtering \cite{SHH07} or wet-chemical synthesis \cite{KT22}, although recent approaches based on atomic-level precision chemical etching have demonstrated nanometer-scale control \cite{PTQ24}.

Here, we employ epitaxy, which provides a robust and reproducible route to synthesize atomically smooth and laterally uniform films, which are crucial for sustaining coherent plasmonic and nonlinear optical responses in a platform suitable for nanophotonic applications. As an indication of the film quality, we clearly observe nearly parabolic electronic bands in angle-resolved photoelectron spectroscopy (Figure~\ref{Fig1}b and Supplementary Figure~S1), associated with vertical QWS in the metallic films \cite{STM06}. The separation between QWS allows us to determine the number of (111) atomic layers in each film, narrowly distributed around a central value; otherwise, in a much thicker (bulk-like) film, the discrete subbands overlap due to inhomogeneous broadening, effectively approaching a continuum-like response. The epitaxial growth mechanism, implemented under carefully controlled pressure and temperature conditions, enables the deposition of high-quality, low-loss crystalline metal films (see Methods). We visualize the presence and density of QWS in Figure~\ref{Fig1}b by comparing angle-resolved photoemission spectroscopy (ARPES) measurements of ultrathin films with two different thicknesses (7 and 14~MLs of silver). We aim to exploit the vertical confinement of conduction electrons in these films to enhance SHG \cite{paper465}: as the film becomes thinner, electronic states are increasingly quantized in the out-of-plane direction, giving rise to more widely separated QWS (Figure~\ref{Fig1}b); this enhanced level spacing produces a more atomic-like optical response and, consequently, stronger nonlinear behavior; in this regime, the out-of-plane electron dynamics departs from the quasi-harmonic behavior characteristic of the bulk metal; equivalently, the QWS no longer resemble the nearly equally spaced ladder states of a parabolic potential.

It is also well established that nonlinearities can be boosted in nanoparticles through the mediation of plasmonic resonances \cite{BBM15,RHV25,MMH16}, which strongly enhance the near field, creating favorable conditions for efficient SHG. To harness this effect, we pattern our epitaxially grown crystalline metal films into periodic nanostructures using EBL, enabling precise control over the lateral geometry (i.e., shape, size, and periodicity; see Supplementary Figure~S2) and the resulting plasmonic resonances. At the plasmon frequency, the electric near field can be substantially enhanced relative to the incident field, as schematically illustrated for a dipolar resonant ribbon in Figure~\ref{Fig1}c.

In this work, we combine these two approaches (i.e., vertical electronic confinement provided by atomically thin crystalline metal films and lateral plasmonic confinement introduced through electron-beam patterning of the same films) to produce a strong second-harmonic response from otherwise centrosymmetric noble-metal crystals. As illustrated schematically in Figure~\ref{Fig1}d, we investigate SHG under different excitation wavelengths from silver ribbons with thicknesses of only a few atomic monolayers (7~MLs of silver capped with 4~MLs of gold in this figure) and lateral widths ranging from tens to hundreds of nanometers. The response of nanoribbon arrays is compared with that of extended films, the substrate signal, and nanotriangles of similar lateral dimensions. The presence of an unpatterned metal film produces a large boost in the SHG signal relative to the bare substrate, while lateral nanostructuring further enhances the nonlinear response, with the highest enhancement found for ribbons. The effect is strongly wavelength-dependent, as illustrated by the presence of additional spectral structure adjacent to the second-harmonic wavelength when the fundamental wavelength is $\lambda_0=1.975~\um$ or $2.3~\um$, in contrast to the sharper response for $\lambda_0=3.1~\um$. In the following, we present a more quantitative study of the SHG generated by atomically thin silver nanoribbons, considering only the SHG signal originating from the region of interest and neglecting any residual contributions observed in the spectra shown in Figure~\ref{Fig1}d.

\begin{figure*}
\centering \includegraphics[width=0.8\textwidth]{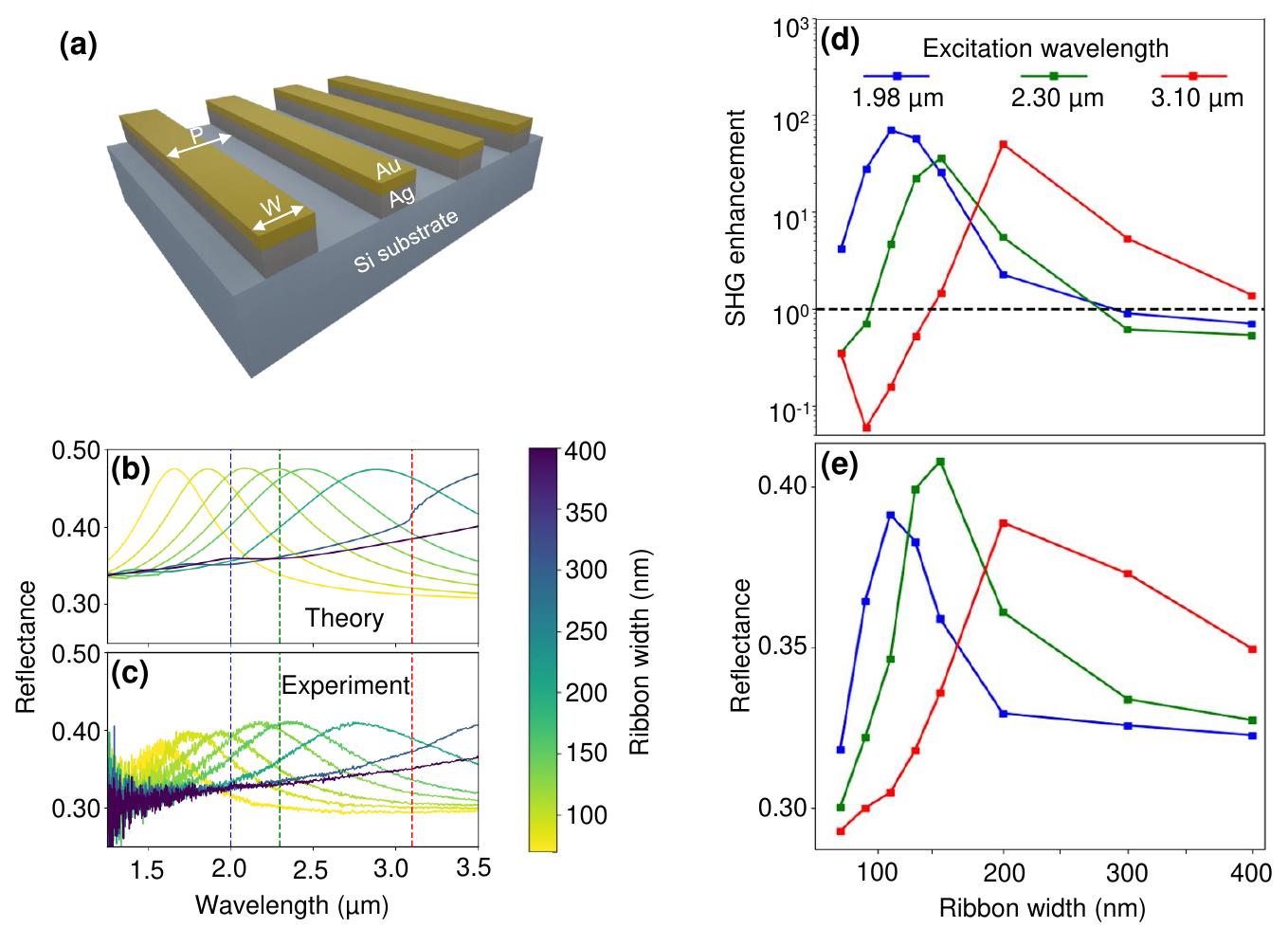}
\caption{{\bf Spectral and geometrical dependence of SHG enhancement.}
\textbf{(a)}~Scheme of the structures under consideration, consisting of Au-capped Ag-ribbon arrays (width $W$, period $P=3\,W$).
\textbf{(b,c)}~Calculated and FTIR-measured reflectance spectra for different ribbon widths.
\textbf{(d)}~Measured SHG enhancement of the ribbon arrays relative to the SHG signal of the homogeneous metal film for three different incident light wavelengths $\lambda_0$ as a function of ribbon width. 
\textbf{(e)}~Reflectance at the three excitation wavelengths for the same sample as in (d).
Silver ribbons lie on a Si substrate, have a fixed thickness of 11\,MLs of Ag(111), and are capped with 4\,MLs of Au(111).}
\label{Fig2}
\end{figure*}

To investigate the enhancement of SHG through engineered plasmonic resonances, we fabricated a set of periodic silver ribbon arrays on a silicon substrate. The period-to-width ratio was set to $P/W=3$ (see Figure~\ref{Fig2}a). In the fabrication process, the width $W$ was varied from 50 to 150~nm in steps of 20~nm, and from 150 to 400~nm in steps of 50~nm. We also examined the thickness dependence of SHG by comparing silver films of 7, 11, and 14~MLs ($\sim1.65~$nm, $\sim2.60$~nm, and $\sim3.31$~nm, respectively), each of them capped with 4~MLs of Au ($\sim1~$nm). Representative secondary-electron-emission (SEM) images are shown in Supplementary Figure~S2.

To determine the spectral position of the plasmonic resonances, we first simulated the reflectance spectra (see Methods) and then validated the results experimentally using Fourier-transform infrared spectroscopy. Measurements were performed over a broad infrared spectral range from $1.2~\um$ to $16~\um$, using a gold mirror as a reference. Simulated and measured reflectance spectra are plotted in Figures~\ref{Fig2}b and \ref{Fig2}c, respectively, for ribbons patterned on 11~ML silver films with increasing widths (colored curves), showing a good overall agreement. The small discrepancies in peak amplitude likely arise from experimental imperfections, including defects in the silver layer, small deviations from the nominal ribbon dimensions, and limitations associated with the EBL fabrication process. We corroborate a scaling of the light wavelength at which the plasmon emerges as $\propto\sqrt{W/d}$ with ribbon width $W$ and film thickness $d$, as predicted in a previous study \cite{paper335}. Such a scaling is maintained for all film thicknesses under consideration (see Supplementary Figure~S3). Based on these results, we correlate the linear optical response with the nonlinear optical measurements, as discussed below.

The resonant plasmonic ribbons significantly enhance the SHG efficiency compared with a planar metal film of the same thickness (see Figure~\ref{Fig2}d). Note that this enhancement originates from the response of the ribbons to polarization along their width, as confirmed by examining the dependence of the SHG signal on the polarization of both the exciting light and the generated output (see Supplementary Figure~S4). In particular, we focus our nonlinear measurements on excitation wavelengths of $1.98~\um$, $2.3~\um$, and $3.1~\um$. The largest enhancements for each excitation wavelength occur when the ribbon width is resonant with the fundamental (excitation) wavelength (cf. Figures~\ref{Fig2}d and \ref{Fig2}e). These results corroborate that near-field enhancement in plasmon-resonant structures increases the SHG signal. Conversely, for ribbon widths that are off-resonance at a given excitation wavelength, the SHG response can be even weaker than that of a planar metal film of the same thickness. We attribute this reduction to the smaller effective lateral interaction area of the ribbons compared with a continuous planar film when no plasmonic enhancement is present. Absolute SHG powers are shown in Supplementary Figure~S5 for $2.3~\um$ excitation as a function of the average pump power. A clear quadratic dependence is observed in the intermediate regime, followed by saturation at sufficiently high powers. Using an analysis analogous to that reported in Ref.~\cite{paper465} for unpatterned films, these results correspond to large effective SHG susceptibilities.

\begin{figure*}
\centering \includegraphics[width=1.0\textwidth]{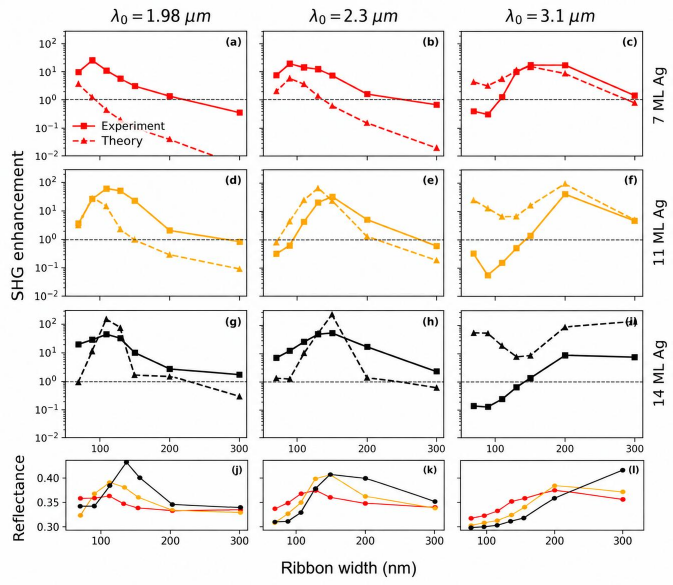}
\caption{{\bf Variation of SHG enhancement with Ag film thicknesses.} (a-i) Experimentally measured (solid curves) and theoretically estimated (dashed curves) SHG enhancement, referenced to the corresponding unpatterned ultrathin metal film, as a function of ribbon width for different film thicknesses (7~MLs (a-c), 11~MLs (d-f), and 14~MLs (g-i) of Ag) and incident light wavelengths ($\lambda_0=1.98~\um$ (a,d,g), $2.3~\um$ (b,e,h), and $3.1~\um$ (c,f,i)). All samples are coated with 4~MLs of Au. (j-l) Measured reflectance at the SHG excitation wavelengths as a function of ribbon width for the thicknesses considered in panels (a-i) (color-coordinated curves) at the same fundamental excitation wavelengths.}
\label{Fig3}
\end{figure*}

We further compare the experimentally measured and theoretically calculated enhancement ratio of the SHG signals produced by ribbon-patterned and unpatterned films for Ag films of three different thicknesses (7, 11, and 14~MLs of silver, from top to bottom, respectively, all coated with 4~MLs of gold) as a function of the ribbon width $W$ and for excitation wavelengths $\lambda_0=1.98~\um$ (left), $2.3~\um$ (center), and $3.1~\um$ (right).

For the experimental data (solid curves), the enhancement factor is obtained by normalizing the peak SHG power measured from the patterned ribbon arrays to that measured from an unpatterned film of the same thickness under identical illumination conditions. The SHG conversion efficiencies of the unpatterned silver films, which serve as the reference for the reported enhancement factors, are shown in Supplementary Figure~S8. The procedure used to extract the SHG conversion efficiencies is illustrated in Supplementary Figure~S5, following the model described in Ref.~\cite{paper465}.

In the theoretical calculations, owing to the centrosymmetric crystallographic structure of the bulk metal, we describe the SHG signal in terms of the nonlinear surface susceptibility, which is assumed to be local, independent of crystallographic orientation, dominated by the normal component $\chi^{\rm SHG}_{\perp\perp\perp}$, and unaffected by film nanostructuring. Under these assumptions, the susceptibility enters as an overall multiplicative factor in the calculation of the far-field SHG amplitude and, consequently, cancels when taking the ratio of the nonlinear intensities generated by patterned and unpatterned films. The induced second-harmonic surface polarization density is given by the product of the nonlinear susceptibility and the square of the normal component of the fundamental near field, and is therefore also directed normal to the surface. By virtue of reciprocity, the far-field amplitude generated by this polarization density at the detector position can be evaluated from the surface-normal field induced by a plane wave incident from the detector direction at the second-harmonic frequency. The nonlinear signal is thus obtained from the surface integral of the squared fundamental surface-normal near field multiplied by the reciprocity-related second-harmonic surface-normal near field (see Methods). Given the in-plane translational invariance of the samples under the illumination conditions considered here, this surface integral reduces to a contour integral (see Supplementary Figure~S10).

The dependence of the SHG enhancement on ribbon width is in qualitative agreement with theory for all excitation wavelengths and film thicknesses considered. The enhancement is peaked at wavelength- and thickness-dependent ribbon widths that are also consistent with the resonances identified from the reflectance of the structured films (Figure~\ref{Fig3}j-l). The SHG enhancement exhibits a significantly more pronounced modulation than the FTIR reflectance, as expected from the nonlinear nature of the process. In addition, the finite spectral bandwidth associated with the pulsed excitation can be neglected relative to the spectral width of the underlying plasmonic resonances. Some discrepancies in magnitude are observed between the experimental and theoretical enhancement ratios. These discrepancies are generally smaller than one order of magnitude, except for the thinnest film thickness at the shortest wavelength. We attribute them primarily to simplifications in the theoretical model, which ignores nonlocal effects and assumes a uniform nonlinear susceptibility that is independent of crystallographic orientation and unaffected by the strong surface-curvature variations at the ribbon edges. Additional effects not included in the model may also contribute, including quantum finite-size effects that modify the surface response, particularly at the smallest thicknesses, as well as imperfections in the fabrication process.

Overall, the observed SHG behavior is consistent with that expected from localized plasmon resonances, as the highest enhancements are observed where the linear reflectance is peaked (Figure~\ref{Fig3}j-l), corresponding to ribbons that are resonant at the respective values of the incident wavelength $\lambda_0$, undergoing plasmon-enhanced local-field confinement. As the ribbon width deviates from the resonance condition, the SHG signal gradually decreases due to the reduced overlap between the fundamental mode and the plasmonic resonance. This behavior highlights the critical role of geometrical tuning in achieving optimal nonlinear conversion. By systematically engineering the ribbon width, SHG can be selectively enhanced at specific excitation wavelengths, providing a versatile degree of tunability in plasmon-enhanced nonlinear structures. The present results should stimulate quantitative predictions of the SHG yield from full electronic-structure calculations that account for the surrounding environment and track the many-body dynamics of thin crystalline metallic structures under strong external illumination.

The SHG enhancement discussed in Figure~\ref{Fig3} focuses on nanoribbon arrays. Comparable and consistent behavior is observed for other geometries, such as periodic nanotriangles (Supplementary Figure~S6), with similar trends after normalization by the metal filling fraction, reflecting geometry-dependent field confinement.

\section{Conclusions}

In summary, we have demonstrated a double mechanism for SHG enhancement from laterally patterned ultrathin crystalline metallic films: on top of the nonlinear enhancement associated with vertical quantum confinement of conduction electrons \cite{paper465}, we observe plasmonic confinement associated with plasmon resonances that produce strong lateral optical-field confinement in patterned nanostructures. We based our results on experimental measurements in good agreement with electromagnetic simulations. Similar levels of SHG enhancement are observed for different types of structures, such as ribbon and triangle arrays, which are compared in Supplementary Figure~S6. A key ingredient in this study has been the stability of the films, consisting of 7-14~MLs of silver passivated with 4~MLs of gold. Looking ahead, lateral engineering of the patterned nanostructures could produce even larger enhancement, including double resonances (i.e., at both the fundamental and the second-harmonic frequencies). Lattice resonances in periodic structures could produce additional boosts of the SHG signal. Other frequency-conversion processes, including spontaneous parametric down-conversion, should also be explored to take advantage of the strong spatial confinement offered by patterned ultrathin metallic films, with potential applications in nanoscale lasers and field amplifiers. Overall, these results offer a promising route toward integration of ultrathin nonlinear plasmonic platforms into sophisticated silicon photonic circuits, enabling on-chip frequency doubling and other nonlinear optical functionalities. Such systems are amenable to miniaturization and tuning, rendering them strong candidates for next-generation nanophotonic devices.

\section*{Methods} \label{sec:Methods}

\subsection*{Fabrication of Atomically Thin Silver Films} Crystalline gold-capped silver films were deposited in an ultra-high vacuum (UHV) chamber with a base pressure of $1.0\times10^{-10}$~mbar on substrates consisting of $4\times12~\text{mm}^2$ n-doped Si(111) chips with a resistivity of $120-340~\Omega~\text{cm}$, corresponding to a dopant concentration of $1.3-3.7\times10^{13}~\text{cm}^{-3}$. This doping concentration ensured a sufficient electrical conductivity for surface science applications and smooth deposition without compromising the plasmonic and optical performance of the Ag-Au hybrid films.

For the processing of the films, the Si(111) substrates were first degassed overnight in UHV at $900~\text{K}$, and then flashed to $1400~\text{K}$ for $20-30~\text{s}$ to remove the native oxide layer. The temperature was subsequently lowered to $600~\text{K}$, maintained for $30~\text{min}$, and then cooled to room temperature, producing an atomically clean, defect-free Si(111) surface with a $7\times7$ reconstruction.  

Silver atoms were evaporated from an electron-bombardment evaporator, with the deposition rate calibrated to sub-monolayer accuracy using a quartz crystal microbalance, and verified by photoemission probing of the characteristic $1-2~\text{ML}$ Ag(111) surface states \cite{SCV05}. Film growth followed a two-step procedure: Ag was deposited with the substrate cooled to $100-120~\text{K}$, and then slowly annealed to room temperature \cite{FMA13}. The typical deposition rate was $\sim0.3~\text{ML/min}$, although comparable quality was obtained for rates between $0.1$ and $0.5~\text{ML/min}$. A few ($\sim4$) MLs of gold were then deposited for film passivation. The deposition temperature, maintained near $100~\text{K}$, was found to be the key parameter for achieving high-quality films. Scanning tunneling microscopy was also used to verify that the surface exhibits sub-atomic-monolayer RMS roughness (Supplementary Figure~S9).

\subsection*{Fabrication of Periodic Nanostructure Arrays} Periodic nanoribbons and nanotriangles of varying sizes were fabricated from these crystalline Ag--Au thin films using a CRESTEC EBL system. The films were first spin-coated with a negative-tone ARN 7520.07 resist at an angular velocity of 4000 revolutions per minute for 1~min, yielding a final resist thickness of $\approx150$~nm. To preserve metal film quality, no baking was applied after spin-coating. The electron-beam source was operated at 50~keV with a current of 50~pA. The area dose of the ARN 7520.07 resist was set to $400~\mu{\rm C}/{\rm cm}^{2}$. An optimum dwell time of $0.7~\us$ was found from a series of test exposures covering the $0.1-1.0$~$\us$ range around the theoretical estimate of $0.72~\us$. These tests were carried out by fixing the write field of the EBL exposure at $60~\um$ with a discretization step of 6~nm. Patterns were exposed in a raster-scanning mode. EBL-exposed regions in the chips were developed by immersion in an AR~300-73 developer for 1.5~min, followed by immersion in DI water for 30~s. The exposed regions resulted in the formation of a patterned resist mask according to the geometrical requirements of the particular nanoparticle arrays. Pattern transfer was carried out by argon plasma dry etching in an Oxford PlasmaLab 80 Plus reactive ion etching (RIE) system, where physical sputtering removed the unprotected Ag--Au thin film regions, thereby yielding the final nanoparticle arrays.

\subsection*{Optical Characterization} Plasmon spectra were acquired in the far field using a Bruker Hyperion Fourier-transform infrared (FTIR) spectrometer operating over a spectral range of $\sim1.2-16~\um$. Measurements were carried out in a reflection geometry under normal incidence with linear polarization across the ribbons, using a gold mirror as a reference for normalization.

\subsection*{SHG Measurements} SHG measurements were conducted using a modified focus-scan setup \cite{paper465} (Supplementary Figure~S7), in which the SHG signal was recorded as the sample was translated along the optical axis through the excitation focus. Lateral positions were probed by moving the tilted sample within the focal plane. The excitation consisted of linearly p-polarized Gaussian pulses (orthogonal to the long axis of the nanoribbons and perpendicular to a side of the equilateral triangles) with pulse duration $\tau \approx 200\,\mathrm{fs}$ and repetition rate $f = 76\,\mathrm{MHz}$, generated by an optical parametric oscillator pumped by a Ti:Sapphire laser. The center wavelength $\lambda_0$ was tunable from 1.975 to $3.1~\um$. Slight daily variations in pulse duration and repetition rate were observed and accounted for.  

A half-wave plate and polarizer ensured linear polarization, while two $20~\mathrm{mm}$ lenses focused the excitation beam (waists $\sigma=11.8$, $13.8$, and $17.0~\um$ characterized for $\lambda_0=1.975$, $2.3$, and $3.1~\um$, respectively, estimated from the intensity $I(x,y)\propto\ee^{-(x^2+y^2)/\sigma^2}$ as a function of lateral position $(x,y)$ relative to the maximum). For the corresponding SHG signals, a reduction of the waist by $\sqrt{2}$ was assumed. Excitation from the back side avoided substrate absorption of the nonlinear signal emitted from the metallic structures placed on the front side. At the silicon interface, the incident angle was refracted from $45^\circ$ in air to $\theta\approx 13^\circ$ inside the substrate, as estimated from Snell's law with the refractive index of the material \cite{FFV18}. After collimation, the nonlinear signal was isolated with transmission filters, coupled into a multimode fiber, and detected by single-photon detectors based on InGaAs or Si ($\lambda_0 = 1.975~\um$). Signal spectra were verified using a Nireos interferometer-based spectrometer via Fourier-transformed autocorrelation. Power drifts during acquisition were below $1\%$ over $20-30~\mathrm{min}$.  

The average excitation power $P_\mathrm{ave}$ was measured before the focusing lens, and the SHG power $P_\mathrm{SHG}$ was extracted from the detector count rate, with both corrected for optical transmission. The SHG response was characterized as a function of excitation power for each region of interest. A quadratic dependence was observed and fitted, while detector dark counts and saturation were included in the model. The SHG conversion efficiency was defined as $\eta = I_{\mathrm{SHG,peak}} / I_{0,\mathrm{peak}}^2$. The excitation peak intensity inside the silver film followed as $I_{0,\mathrm{peak}} = 2 P_\mathrm{peak} / A$, where $P_\mathrm{peak}$ is the peak power of a single pulse and $A$ the illumination area. For oblique incidence, an area $A = \pi\sigma_0^2 \sec\theta$ was approximated for an ellipse defined from the beam waist $\sigma_0$ and incidence angle $\theta$ inside the silicon. The peak power was obtained from the average power as $P_\mathrm{peak}=(f_\mathrm{shape}/f\tau)\, P_\mathrm{ave}$, where the factor $f_\mathrm{shape}=0.94$ corrects for lateral Gaussian pulse profiles.

\subsection*{Theoretical Calculation of Enhancement Factors} We define the enhancement ratio as the SHG intensity produced upon transmission of light through nanostructured ultrathin metal films divided by that produced by extended homogeneous films. For simplicity, we restrict our calculations to ribbon arrays, although the approach can be readily extended to periodic arrays of triangles or other geometries by integrating over the two-dimensional in-plane coordinates.

We assume that the SHG intensity scales with the squared magnitude of the second-harmonic polarization, which, as a rough approximation, is taken to be locally proportional to the square of the linear electric field at the fundamental frequency $\omega$. The electric near fields $\Eb_\omega^{\rm NR}$ and $\Eb_\omega^{\rm TF}$ right outside the surfaces of the metal nanoribbons (NRs) and uniform thin films (TFs), respectively, are obtained numerically, as detailed below. For homogeneous films, the numerical results are in close agreement with the analytical theory for planar interfaces \cite{NH06}.

Because the crystal lattice of the metal is centrosymmetric, we only consider an SHG surface polarization in both the structured and uniform films. This quantity is proportional to the nonlinear surface susceptibility, which we assume to be local, independent of surface orientation, and unaffected by nanostructuring (i.e., identical in the two configurations). We further assume that the surface susceptibility is dominated by the surface-normal component $\chi^{\rm SHG}_{\perp\perp\perp}$. Consequently, the susceptibility enters the SHG far-field intensity as an overall multiplicative factor that cancels when taking the ratio of the nonlinear intensities generated by patterned and unpatterned films.

We thus have a surface polarization oriented perpendicular to the metal boundary, whose magnitude is proportional to the nonlinear susceptibility and to $|E_{\omega,\perp}|^2$. To determine the radiation reaching the detector, we invoke electromagnetic reciprocity rather than explicitly propagating the emitted second-harmonic field. Specifically, the coupling to the detected far-field mode is obtained from the normal electric-field component at $2\omega$ generated when the structure is illuminated from the detection direction. The resulting SHG amplitude can therefore be expressed as an integral over the metal surface involving the product of the local fundamental-field factor $[E_{\omega,\perp}]^2$ and the corresponding reciprocal field at the second-harmonic frequency $E_{2\omega,\perp}$. For the illumination geometry considered here, translational symmetry along the invariant in-plane direction allows the surface integration to be reduced to a one-dimensional contour integral, as illustrated in Supplementary Figure~S10. From these considerations, the enhancement ratio reduces to
\begin{widetext}
\begin{align} \label{SHGE} 
\text{SHG enhancement}=
\left|\dfrac{\displaystyle\oint_{\rm NR}\!\!\!ds\;\; \big[E_{\omega,\perp}^{\rm NR}(\rb_s)\big]^2\, E_{2\omega,\perp}^{\rm NR}(\rb_s)}{\displaystyle\int_{\rm TF}\!\!\!ds\;\; \big[E_{\omega,\perp}^{\rm TF}(\rb_s)\big]^2\, E_{2\omega,\perp}^{\rm TF}(\rb_s)}\right|^2,
\end{align}
\end{widetext}
where the integrations are performed along the NR and TF boundaries, respectively. For the NR, the path forms a closed loop around a single ribbon, whereas in the TF it follows the upper and lower interfaces over a distance equal to the period of the corresponding patterned structure. We use $s$ as the arc-length parameter along the respective contours. This model is used to generate the theoretical results shown in Figure~\ref{Fig3}.

\subsection*{Simulation of Optical Electric Fields} Near fields in the nanostructures were calculated from 2D simulations in the frequency domain using COMSOL Multiphysics, introducing periodic boundary conditions along the lateral direction of array symmetry to emulate the periodic structure, and perfectly matched layers along the out-of-plane direction. We considered a p-polarized incident plane wave impinging on the air--Si interface at an angle of $45^\circ$ and in-plane orientation as in the experiment. A Drude-like permittivity $\epsilon_m(\omega)=\epsilon_b-\omega_p^2/[\omega(\omega+\ii\xi\gamma)]$ was used for the metals, with parameters $\hbar\omega_p=9.17$~eV, $\hbar\gamma=21$~meV, and $\epsilon_b=4$ for silver, and $\hbar\omega_p=9.06$~eV, $\hbar\gamma=71$~meV, and $\epsilon_b=9.5$ in gold, as extracted by fitting measured optical data \cite{JC1972}. A correction factor $\xi$ was introduced to account for additional damping produced by imperfections introduced during the fabrication process (see main text). More precisely, results are fitted with $\xi=3$, 5, and 7 for films containing 14, 11, and 7 Ag MLs, respectively.

\section*{Acknowledgments}
\noindent This work has been supported in part by the European Research Council (101141220-QUEFES), the European Commission (101135288-EPIQUE), the Spanish MICIU (PID2024-157421NB-I00 and Severo Ochoa CEX2024-001490-S), the CERCA program, the Basque Gouvernment (IT2133-26), and the Austrian Science Fund (10.55776/COE1, 10.55776/F71, and 10.55776/FG5).


\begin{thebibliography}{47}%
\makeatletter
\providecommand \@ifxundefined [1]{%
 \@ifx{#1\undefined}
}%
\providecommand \@ifnum [1]{%
 \ifnum #1\expandafter \@firstoftwo
 \else \expandafter \@secondoftwo
 \fi
}%
\providecommand \@ifx [1]{%
 \ifx #1\expandafter \@firstoftwo
 \else \expandafter \@secondoftwo
 \fi
}%
\providecommand \natexlab [1]{#1}%
\providecommand \enquote  [1]{``#1''}%
\providecommand \bibnamefont  [1]{#1}%
\providecommand \bibfnamefont [1]{#1}%
\providecommand \citenamefont [1]{#1}%
\providecommand \href@noop [0]{\@secondoftwo}%
\providecommand \href [0]{\begingroup \@sanitize@url \@href}%
\providecommand \@href[1]{\@@startlink{#1}\@@href}%
\providecommand \@@href[1]{\endgroup#1\@@endlink}%
\providecommand \@sanitize@url [0]{\catcode `\\12\catcode `\$12\catcode
  `\&12\catcode `\#12\catcode `\^12\catcode `\_12\catcode `\%12\relax}%
\providecommand \@@startlink[1]{}%
\providecommand \@@endlink[0]{}%
\providecommand \url  [0]{\begingroup\@sanitize@url \@url }%
\providecommand \@url [1]{\endgroup\@href {#1}{\urlprefix }}%
\providecommand \urlprefix  [0]{URL }%
\providecommand \Eprint [0]{\href }%
\providecommand \doibase [0]{http://dx.doi.org/}%
\providecommand \selectlanguage [0]{\@gobble}%
\providecommand \bibinfo  [0]{\@secondoftwo}%
\providecommand \bibfield  [0]{\@secondoftwo}%
\providecommand \translation [1]{[#1]}%
\providecommand \BibitemOpen [0]{}%
\providecommand \bibitemStop [0]{}%
\providecommand \bibitemNoStop [0]{.\EOS\space}%
\providecommand \EOS [0]{\spacefactor3000\relax}%
\providecommand \BibitemShut  [1]{\csname bibitem#1\endcsname}%
\let\auto@bib@innerbib\@empty
\bibitem [{\citenamefont {Sain}\ \emph {et~al.}(2019)\citenamefont {Sain},
  \citenamefont {Meier},\ and\ \citenamefont {Zentgraf}}]{SMZ19}%
  \BibitemOpen
  \bibfield  {author} {\bibinfo {author} {\bibfnamefont {B.}~\bibnamefont
  {Sain}}, \bibinfo {author} {\bibfnamefont {C.}~\bibnamefont {Meier}}, \ and\
  \bibinfo {author} {\bibfnamefont {T.}~\bibnamefont {Zentgraf}},\ }\bibfield
  {title} {\enquote {\bibinfo {title} {Nonlinear optics in all-dielectric
  nanoantennas and metasurfaces: a review},}\ }\href {\doibase
  10.1117/1.AP.1.2.024002} {\bibfield  {journal} {\bibinfo  {journal} {Adv.\
  Photonics}\ }\textbf {\bibinfo {volume} {1}},\ \bibinfo {pages} {024002}
  (\bibinfo {year} {2019})}\BibitemShut {NoStop}%
\bibitem [{\citenamefont {McKenna}\ \emph {et~al.}(2022)\citenamefont
  {McKenna}, \citenamefont {Stokowski}, \citenamefont {Ansari}, \citenamefont
  {Mishra}, \citenamefont {Jankowski}, \citenamefont {Sarabalis}, \citenamefont
  {Herrmann}, \citenamefont {Langrock}, \citenamefont {Fejer},\ and\
  \citenamefont {Safavi-Naeini}}]{MSA22}%
  \BibitemOpen
  \bibfield  {author} {\bibinfo {author} {\bibfnamefont {T.~P.}\ \bibnamefont
  {McKenna}}, \bibinfo {author} {\bibfnamefont {H.~S.}\ \bibnamefont
  {Stokowski}}, \bibinfo {author} {\bibfnamefont {V.}~\bibnamefont {Ansari}},
  \bibinfo {author} {\bibfnamefont {J.}~\bibnamefont {Mishra}}, \bibinfo
  {author} {\bibfnamefont {M.}~\bibnamefont {Jankowski}}, \bibinfo {author}
  {\bibfnamefont {C.~J.}\ \bibnamefont {Sarabalis}}, \bibinfo {author}
  {\bibfnamefont {J.~F.}\ \bibnamefont {Herrmann}}, \bibinfo {author}
  {\bibfnamefont {C.}~\bibnamefont {Langrock}}, \bibinfo {author}
  {\bibfnamefont {M.~M.}\ \bibnamefont {Fejer}}, \ and\ \bibinfo {author}
  {\bibfnamefont {A.~H.}\ \bibnamefont {Safavi-Naeini}},\ }\bibfield  {title}
  {\enquote {\bibinfo {title} {Ultra-low-power second-order nonlinear optics on
  a chip},}\ }\href {\doibase 10.1038/s41467-022-31134-5} {\bibfield  {journal}
  {\bibinfo  {journal} {Nat.\ Commun.}\ }\textbf {\bibinfo {volume} {13}},\
  \bibinfo {pages} {4532} (\bibinfo {year} {2022})}\BibitemShut {NoStop}%
\bibitem [{\citenamefont {Adeshina}\ and\ \citenamefont {Kim}(2025)}]{AK25}%
  \BibitemOpen
  \bibfield  {author} {\bibinfo {author} {\bibfnamefont {M.~A.}\ \bibnamefont
  {Adeshina}}\ and\ \bibinfo {author} {\bibfnamefont {H.}~\bibnamefont {Kim}},\
  }\bibfield  {title} {\enquote {\bibinfo {title} {Exploring the frontier:
  nonlinear optics in low dimensional materials},}\ }\href {\doibase
  10.1515/nanoph-2024-0652} {\bibfield  {journal} {\bibinfo  {journal}
  {Nanophotonics}\ }\textbf {\bibinfo {volume} {14}},\ \bibinfo {pages}
  {1451--1473} (\bibinfo {year} {2025})}\BibitemShut {NoStop}%
\bibitem [{\citenamefont {Luo}\ \emph {et~al.}(2018)\citenamefont {Luo},
  \citenamefont {He}, \citenamefont {Liang}, \citenamefont {Li},\ and\
  \citenamefont {Lin}}]{LHL18}%
  \BibitemOpen
  \bibfield  {author} {\bibinfo {author} {\bibfnamefont {R.}~\bibnamefont
  {Luo}}, \bibinfo {author} {\bibfnamefont {Y.}~\bibnamefont {He}}, \bibinfo
  {author} {\bibfnamefont {H.}~\bibnamefont {Liang}}, \bibinfo {author}
  {\bibfnamefont {M.}~\bibnamefont {Li}}, \ and\ \bibinfo {author}
  {\bibfnamefont {Q.}~\bibnamefont {Lin}},\ }\bibfield  {title} {\enquote
  {\bibinfo {title} {Highly tunable efficient second-harmonic generation in a
  lithium niobate nanophotonic waveguide},}\ }\href {\doibase
  10.1364/OPTICA.5.001006} {\bibfield  {journal} {\bibinfo  {journal} {Optica}\
  }\textbf {\bibinfo {volume} {5}},\ \bibinfo {pages} {1006--1011} (\bibinfo
  {year} {2018})}\BibitemShut {NoStop}%
\bibitem [{\citenamefont {Qu}\ \emph {et~al.}(2025)\citenamefont {Qu},
  \citenamefont {Wu}, \citenamefont {Cai}, \citenamefont {Ren},\ and\
  \citenamefont {Xu}}]{QWC25}%
  \BibitemOpen
  \bibfield  {author} {\bibinfo {author} {\bibfnamefont {L.}~\bibnamefont
  {Qu}}, \bibinfo {author} {\bibfnamefont {W.}~\bibnamefont {Wu}}, \bibinfo
  {author} {\bibfnamefont {W.}~\bibnamefont {Cai}}, \bibinfo {author}
  {\bibfnamefont {M.}~\bibnamefont {Ren}}, \ and\ \bibinfo {author}
  {\bibfnamefont {J.}~\bibnamefont {Xu}},\ }\bibfield  {title} {\enquote
  {\bibinfo {title} {Second harmonic generation in lithium niobate on
  insulator},}\ }\href {\doibase 10.1002/lpor.202401928} {\bibfield  {journal}
  {\bibinfo  {journal} {Laser\ Photonics\ Rev.}\ }\textbf {\bibinfo {volume}
  {19}},\ \bibinfo {pages} {2401928} (\bibinfo {year} {2025})}\BibitemShut
  {NoStop}%
\bibitem [{\citenamefont {Yamamoto}\ \emph {et~al.}(1995)\citenamefont
  {Yamamoto}, \citenamefont {Yamaguchi}, \citenamefont {Yamada}, \citenamefont
  {Ueda},\ and\ \citenamefont {Matsumoto}}]{YYY95}%
  \BibitemOpen
  \bibfield  {author} {\bibinfo {author} {\bibfnamefont {Y.}~\bibnamefont
  {Yamamoto}}, \bibinfo {author} {\bibfnamefont {S.}~\bibnamefont {Yamaguchi}},
  \bibinfo {author} {\bibfnamefont {N.}~\bibnamefont {Yamada}}, \bibinfo
  {author} {\bibfnamefont {K.}~\bibnamefont {Ueda}}, \ and\ \bibinfo {author}
  {\bibfnamefont {T.}~\bibnamefont {Matsumoto}},\ }\bibfield  {title} {\enquote
  {\bibinfo {title} {Quasi-phase-matched second-harmonic generation in a
  periodic-lens sequence waveguide with a relatively wide wavelength-tuned
  width},}\ }\href {\doibase 10.1143/JJAP.34.6382} {\bibfield  {journal}
  {\bibinfo  {journal} {Jpn.\ J.\ Appl.\ Phys.}\ }\textbf {\bibinfo {volume}
  {34}},\ \bibinfo {pages} {6382} (\bibinfo {year} {1995})}\BibitemShut
  {NoStop}%
\bibitem [{\citenamefont {Boyd}(2008)}]{B08_3}%
  \BibitemOpen
  \bibfield  {author} {\bibinfo {author} {\bibfnamefont {R.~W.}\ \bibnamefont
  {Boyd}},\ }\href@noop {} {\emph {\bibinfo {title} {Nonlinear Optics}}},\
  \bibinfo {edition} {3rd}\ ed.\ (\bibinfo  {publisher} {Academic Press},\
  \bibinfo {address} {Amsterdam},\ \bibinfo {year} {2008})\BibitemShut
  {NoStop}%
\bibitem [{\citenamefont {Kauranen}\ and\ \citenamefont {Zayats}(2012)}]{KZ12}%
  \BibitemOpen
  \bibfield  {author} {\bibinfo {author} {\bibfnamefont {M.}~\bibnamefont
  {Kauranen}}\ and\ \bibinfo {author} {\bibfnamefont {A.~V.}\ \bibnamefont
  {Zayats}},\ }\bibfield  {title} {\enquote {\bibinfo {title} {Nonlinear
  plasmonics},}\ }\href {\doibase 10.1038/nphoton.2012.244} {\bibfield
  {journal} {\bibinfo  {journal} {Nat.\ Photonics}\ }\textbf {\bibinfo {volume}
  {6}},\ \bibinfo {pages} {737--748} (\bibinfo {year} {2012})}\BibitemShut
  {NoStop}%
\bibitem [{\citenamefont {Liang}\ \emph {et~al.}(2014)\citenamefont {Liang},
  \citenamefont {Sun}, \citenamefont {Jiang}, \citenamefont {Jiang},\ and\
  \citenamefont {Chen}}]{LSJ14}%
  \BibitemOpen
  \bibfield  {author} {\bibinfo {author} {\bibfnamefont {Z.}~\bibnamefont
  {Liang}}, \bibinfo {author} {\bibfnamefont {J.}~\bibnamefont {Sun}}, \bibinfo
  {author} {\bibfnamefont {Y.}~\bibnamefont {Jiang}}, \bibinfo {author}
  {\bibfnamefont {L.}~\bibnamefont {Jiang}}, \ and\ \bibinfo {author}
  {\bibfnamefont {X.}~\bibnamefont {Chen}},\ }\bibfield  {title} {\enquote
  {\bibinfo {title} {Plasmonic enhanced optoelectronic devices},}\ }\href
  {\doibase 10.1007/s11468-014-9682-7} {\bibfield  {journal} {\bibinfo
  {journal} {Plasmonics}\ }\textbf {\bibinfo {volume} {9}},\ \bibinfo {pages}
  {859--866} (\bibinfo {year} {2014})}\BibitemShut {NoStop}%
\bibitem [{\citenamefont {Wu}\ \emph {et~al.}(2021)\citenamefont {Wu},
  \citenamefont {Jiang}, \citenamefont {Wang}, \citenamefont {Zhao},
  \citenamefont {Shi}, \citenamefont {Zhang}, \citenamefont {Chen},
  \citenamefont {Zhang}, \citenamefont {Zhang},\ and\ \citenamefont
  {Liu}}]{WJW21}%
  \BibitemOpen
  \bibfield  {author} {\bibinfo {author} {\bibfnamefont {X.-X.}\ \bibnamefont
  {Wu}}, \bibinfo {author} {\bibfnamefont {W.-Y.}\ \bibnamefont {Jiang}},
  \bibinfo {author} {\bibfnamefont {X.-F.}\ \bibnamefont {Wang}}, \bibinfo
  {author} {\bibfnamefont {L.-Y.}\ \bibnamefont {Zhao}}, \bibinfo {author}
  {\bibfnamefont {J.}~\bibnamefont {Shi}}, \bibinfo {author} {\bibfnamefont
  {S.}~\bibnamefont {Zhang}}, \bibinfo {author} {\bibfnamefont {Z.-X.}\
  \bibnamefont {Chen}}, \bibinfo {author} {\bibfnamefont {W.}~\bibnamefont
  {Zhang}}, \bibinfo {author} {\bibfnamefont {Y.}~\bibnamefont {Zhang}}, \ and\
  \bibinfo {author} {\bibfnamefont {X.-F.}\ \bibnamefont {Liu}},\ }\bibfield
  {title} {\enquote {\bibinfo {title} {Inch-scale ball-in-bowl plasmonic
  nanostructure arrays for polarization-independent second-harmonic
  generation},}\ }\href {\doibase 10.1021/acsnano.0c08498} {\bibfield
  {journal} {\bibinfo  {journal} {ACS\ Nano}\ }\textbf {\bibinfo {volume}
  {15}},\ \bibinfo {pages} {1291--1300} (\bibinfo {year} {2021})}\BibitemShut
  {NoStop}%
\bibitem [{\citenamefont {Lu}\ \emph {et~al.}(2023)\citenamefont {Lu},
  \citenamefont {Luo}, \citenamefont {Gao}, \citenamefont {Wang}, \citenamefont
  {Sun},\ and\ \citenamefont {Nam}}]{LLG23}%
  \BibitemOpen
  \bibfield  {author} {\bibinfo {author} {\bibfnamefont {K.}~\bibnamefont
  {Lu}}, \bibinfo {author} {\bibfnamefont {M.}~\bibnamefont {Luo}}, \bibinfo
  {author} {\bibfnamefont {W.}~\bibnamefont {Gao}}, \bibinfo {author}
  {\bibfnamefont {Q.~J.}\ \bibnamefont {Wang}}, \bibinfo {author}
  {\bibfnamefont {H.}~\bibnamefont {Sun}}, \ and\ \bibinfo {author}
  {\bibfnamefont {D.}~\bibnamefont {Nam}},\ }\bibfield  {title} {\enquote
  {\bibinfo {title} {Strong second-harmonic generation by sublattice
  polarization in non-uniformly strained monolayer graphene},}\ }\href
  {\doibase 10.1038/s41467-023-38344-5} {\bibfield  {journal} {\bibinfo
  {journal} {Nat.\ Commun.}\ }\textbf {\bibinfo {volume} {14}},\ \bibinfo
  {pages} {2580} (\bibinfo {year} {2023})}\BibitemShut {NoStop}%
\bibitem [{\citenamefont {Kim}\ \emph {et~al.}(2019)\citenamefont {Kim},
  \citenamefont {Fr{\"o}ch}, \citenamefont {Gardner}, \citenamefont {Li},
  \citenamefont {Aharonovich},\ and\ \citenamefont {Solntsev}}]{KFG19}%
  \BibitemOpen
  \bibfield  {author} {\bibinfo {author} {\bibfnamefont {S.}~\bibnamefont
  {Kim}}, \bibinfo {author} {\bibfnamefont {J.~E.}\ \bibnamefont {Fr{\"o}ch}},
  \bibinfo {author} {\bibfnamefont {A.}~\bibnamefont {Gardner}}, \bibinfo
  {author} {\bibfnamefont {C.}~\bibnamefont {Li}}, \bibinfo {author}
  {\bibfnamefont {I.}~\bibnamefont {Aharonovich}}, \ and\ \bibinfo {author}
  {\bibfnamefont {A.~S.}\ \bibnamefont {Solntsev}},\ }\bibfield  {title}
  {\enquote {\bibinfo {title} {Second-harmonic generation in multilayer
  hexagonal boron nitride flakes},}\ }\href {\doibase 10.1364/OL.44.005792}
  {\bibfield  {journal} {\bibinfo  {journal} {Opt.\ Lett.}\ }\textbf {\bibinfo
  {volume} {44}},\ \bibinfo {pages} {5792--5795} (\bibinfo {year}
  {2019})}\BibitemShut {NoStop}%
\bibitem [{\citenamefont {Trovatello}\ \emph {et~al.}(2025)\citenamefont
  {Trovatello}, \citenamefont {Ferrante}, \citenamefont {Yang}, \citenamefont
  {Bajo}, \citenamefont {Braun}, \citenamefont {Peng}, \citenamefont {Xu},
  \citenamefont {Jenke}, \citenamefont {Ye}, \citenamefont {Delor},
  \citenamefont {Basov}, \citenamefont {Park}, \citenamefont {Walther},
  \citenamefont {Dean}, \citenamefont {Rozema}, \citenamefont {Marini},
  \citenamefont {Cerullo},\ and\ \citenamefont {Schuck}}]{TFY25}%
  \BibitemOpen
  \bibfield  {author} {\bibinfo {author} {\bibfnamefont {C.}~\bibnamefont
  {Trovatello}}, \bibinfo {author} {\bibfnamefont {C.}~\bibnamefont
  {Ferrante}}, \bibinfo {author} {\bibfnamefont {B.}~\bibnamefont {Yang}},
  \bibinfo {author} {\bibfnamefont {J.}~\bibnamefont {Bajo}}, \bibinfo {author}
  {\bibfnamefont {B.}~\bibnamefont {Braun}}, \bibinfo {author} {\bibfnamefont
  {Z.~H.}\ \bibnamefont {Peng}}, \bibinfo {author} {\bibfnamefont
  {X.}~\bibnamefont {Xu}}, \bibinfo {author} {\bibfnamefont {Ph.~K.}\
  \bibnamefont {Jenke}}, \bibinfo {author} {\bibfnamefont {A.}~\bibnamefont
  {Ye}}, \bibinfo {author} {\bibfnamefont {M.}~\bibnamefont {Delor}}, \bibinfo
  {author} {\bibfnamefont {D.~N.}\ \bibnamefont {Basov}}, \bibinfo {author}
  {\bibfnamefont {J.}~\bibnamefont {Park}}, \bibinfo {author} {\bibfnamefont
  {Ph.}\ \bibnamefont {Walther}}, \bibinfo {author} {\bibfnamefont {C.~R.}\
  \bibnamefont {Dean}}, \bibinfo {author} {\bibfnamefont {L.~A.}\ \bibnamefont
  {Rozema}}, \bibinfo {author} {\bibfnamefont {A.}~\bibnamefont {Marini}},
  \bibinfo {author} {\bibfnamefont {G.}~\bibnamefont {Cerullo}}, \ and\
  \bibinfo {author} {\bibfnamefont {P.~J.}\ \bibnamefont {Schuck}},\ }\bibfield
   {title} {\enquote {\bibinfo {title} {Quasi-phase-matched up- and
  down-conversion in periodically poled layered semiconductors},}\ }\href
  {\doibase 10.1038/s41566-024-01602-z} {\bibfield  {journal} {\bibinfo
  {journal} {Nat.\ Photonics}\ }\textbf {\bibinfo {volume} {19}},\ \bibinfo
  {pages} {291--299} (\bibinfo {year} {2025})}\BibitemShut {NoStop}%
\bibitem [{\citenamefont {Zielinski}\ \emph {et~al.}(2009)\citenamefont
  {Zielinski}, \citenamefont {Oron}, \citenamefont {Chauvat},\ and\
  \citenamefont {Zyss}}]{ZOC09}%
  \BibitemOpen
  \bibfield  {author} {\bibinfo {author} {\bibfnamefont {M.}~\bibnamefont
  {Zielinski}}, \bibinfo {author} {\bibfnamefont {D.}~\bibnamefont {Oron}},
  \bibinfo {author} {\bibfnamefont {D.}~\bibnamefont {Chauvat}}, \ and\
  \bibinfo {author} {\bibfnamefont {J.}~\bibnamefont {Zyss}},\ }\bibfield
  {title} {\enquote {\bibinfo {title} {Second-harmonic generation from a single
  core/shell quantum dot},}\ }\href {\doibase 10.1002/smll.200900399}
  {\bibfield  {journal} {\bibinfo  {journal} {Small}\ }\textbf {\bibinfo
  {volume} {5}},\ \bibinfo {pages} {2835--2840} (\bibinfo {year}
  {2009})}\BibitemShut {NoStop}%
\bibitem [{\citenamefont {Ren}\ \emph {et~al.}(2021)\citenamefont {Ren},
  \citenamefont {Chen}, \citenamefont {Li}, \citenamefont {Wang}, \citenamefont
  {Zhao}, \citenamefont {Zhao}, \citenamefont {Hu}, \citenamefont {Qu},\ and\
  \citenamefont {Liu}}]{RCL21}%
  \BibitemOpen
  \bibfield  {author} {\bibinfo {author} {\bibfnamefont {S.}~\bibnamefont
  {Ren}}, \bibinfo {author} {\bibfnamefont {Z.}~\bibnamefont {Chen}}, \bibinfo
  {author} {\bibfnamefont {S.}~\bibnamefont {Li}}, \bibinfo {author}
  {\bibfnamefont {S.}~\bibnamefont {Wang}}, \bibinfo {author} {\bibfnamefont
  {Z.}~\bibnamefont {Zhao}}, \bibinfo {author} {\bibfnamefont {Y.}~\bibnamefont
  {Zhao}}, \bibinfo {author} {\bibfnamefont {R.}~\bibnamefont {Hu}}, \bibinfo
  {author} {\bibfnamefont {J.}~\bibnamefont {Qu}}, \ and\ \bibinfo {author}
  {\bibfnamefont {L.}~\bibnamefont {Liu}},\ }\bibfield  {title} {\enquote
  {\bibinfo {title} {Resonance-enhanced second harmonic generation via quantum
  dots integrated with ag nanoarrays},}\ }\href {\doibase 10.1364/OME.436096}
  {\bibfield  {journal} {\bibinfo  {journal} {Opt.\ Mater.\ Express}\ }\textbf
  {\bibinfo {volume} {11}},\ \bibinfo {pages} {3223--3231} (\bibinfo {year}
  {2021})}\BibitemShut {NoStop}%
\bibitem [{\citenamefont {Wang}\ \emph {et~al.}(2009)\citenamefont {Wang},
  \citenamefont {Rodr\'{\i}guez}, \citenamefont {Albers}, \citenamefont
  {Ahorinta}, \citenamefont {Sipe},\ and\ \citenamefont {Kauranen}}]{WRA09}%
  \BibitemOpen
  \bibfield  {author} {\bibinfo {author} {\bibfnamefont {F.~X.}\ \bibnamefont
  {Wang}}, \bibinfo {author} {\bibfnamefont {F.~J.}\ \bibnamefont
  {Rodr\'{\i}guez}}, \bibinfo {author} {\bibfnamefont {W.~M.}\ \bibnamefont
  {Albers}}, \bibinfo {author} {\bibfnamefont {R.}~\bibnamefont {Ahorinta}},
  \bibinfo {author} {\bibfnamefont {J.~E.}\ \bibnamefont {Sipe}}, \ and\
  \bibinfo {author} {\bibfnamefont {M.}~\bibnamefont {Kauranen}},\ }\bibfield
  {title} {\enquote {\bibinfo {title} {Surface and bulk contributions to the
  second-order nonlinear optical response of a gold film},}\ }\href {\doibase
  10.1103/PhysRevB.80.233402} {\bibfield  {journal} {\bibinfo  {journal}
  {Phys.\ Rev.\ B}\ }\textbf {\bibinfo {volume} {80}},\ \bibinfo {pages}
  {233402} (\bibinfo {year} {2009})}\BibitemShut {NoStop}%
\bibitem [{\citenamefont {Boyd}\ \emph {et~al.}(2014)\citenamefont {Boyd},
  \citenamefont {Shi},\ and\ \citenamefont {{De Leon}}}]{BSD14}%
  \BibitemOpen
  \bibfield  {author} {\bibinfo {author} {\bibfnamefont {R.~W.}\ \bibnamefont
  {Boyd}}, \bibinfo {author} {\bibfnamefont {Z.}~\bibnamefont {Shi}}, \ and\
  \bibinfo {author} {\bibfnamefont {I.}~\bibnamefont {{De Leon}}},\ }\bibfield
  {title} {\enquote {\bibinfo {title} {The third-order nonlinear optical
  susceptibility of gold},}\ }\href {\doibase 10.1016/j.optcom.2014.03.005}
  {\bibfield  {journal} {\bibinfo  {journal} {Opt.\ Commun.}\ }\textbf
  {\bibinfo {volume} {326}},\ \bibinfo {pages} {74--79} (\bibinfo {year}
  {2014})}\BibitemShut {NoStop}%
\bibitem [{\citenamefont {Dai}\ \emph {et~al.}(2017)\citenamefont {Dai},
  \citenamefont {Zhang}, \citenamefont {Wang}, \citenamefont {Wang},
  \citenamefont {Zhang}, \citenamefont {Gong}, \citenamefont {Han},\ and\
  \citenamefont {Han}}]{DZW17}%
  \BibitemOpen
  \bibfield  {author} {\bibinfo {author} {\bibfnamefont {H.}~\bibnamefont
  {Dai}}, \bibinfo {author} {\bibfnamefont {L.}~\bibnamefont {Zhang}}, \bibinfo
  {author} {\bibfnamefont {Z.}~\bibnamefont {Wang}}, \bibinfo {author}
  {\bibfnamefont {X.}~\bibnamefont {Wang}}, \bibinfo {author} {\bibfnamefont
  {J.}~\bibnamefont {Zhang}}, \bibinfo {author} {\bibfnamefont
  {H.}~\bibnamefont {Gong}}, \bibinfo {author} {\bibfnamefont {J.-B.}\
  \bibnamefont {Han}}, \ and\ \bibinfo {author} {\bibfnamefont
  {Y.}~\bibnamefont {Han}},\ }\bibfield  {title} {\enquote {\bibinfo {title}
  {Linear and nonlinear optical properties of silver-coated gold nanorods},}\
  }\href {\doibase 10.1021/acs.jpcc.7b00295} {\bibfield  {journal} {\bibinfo
  {journal} {J.\ Phys.\ Chem.\ C}\ }\textbf {\bibinfo {volume} {121}},\
  \bibinfo {pages} {12358--12364} (\bibinfo {year} {2017})}\BibitemShut
  {NoStop}%
\bibitem [{\citenamefont {Abdelwahab}\ \emph {et~al.}(2022)\citenamefont
  {Abdelwahab}, \citenamefont {Tilmann}, \citenamefont {Wu}, \citenamefont
  {Giovanni}, \citenamefont {Verzhbitskiy}, \citenamefont {Zhu}, \citenamefont
  {Bert{\'e}}, \citenamefont {Xuan}, \citenamefont {de~S.~Menezes},
  \citenamefont {Eda}, \citenamefont {Sum}, \citenamefont {Quek}, \citenamefont
  {Maier},\ and\ \citenamefont {Loh}}]{ATW22}%
  \BibitemOpen
  \bibfield  {author} {\bibinfo {author} {\bibfnamefont {I.}~\bibnamefont
  {Abdelwahab}}, \bibinfo {author} {\bibfnamefont {B.}~\bibnamefont {Tilmann}},
  \bibinfo {author} {\bibfnamefont {Y.}~\bibnamefont {Wu}}, \bibinfo {author}
  {\bibfnamefont {D.}~\bibnamefont {Giovanni}}, \bibinfo {author}
  {\bibfnamefont {I.}~\bibnamefont {Verzhbitskiy}}, \bibinfo {author}
  {\bibfnamefont {M.}~\bibnamefont {Zhu}}, \bibinfo {author} {\bibfnamefont
  {R.}~\bibnamefont {Bert{\'e}}}, \bibinfo {author} {\bibfnamefont
  {F.}~\bibnamefont {Xuan}}, \bibinfo {author} {\bibfnamefont {L.}~\bibnamefont
  {de~S.~Menezes}}, \bibinfo {author} {\bibfnamefont {G.}~\bibnamefont {Eda}},
  \bibinfo {author} {\bibfnamefont {T.~C.}\ \bibnamefont {Sum}}, \bibinfo
  {author} {\bibfnamefont {S.~Y.}\ \bibnamefont {Quek}}, \bibinfo {author}
  {\bibfnamefont {S.~A.}\ \bibnamefont {Maier}}, \ and\ \bibinfo {author}
  {\bibfnamefont {K.~P.}\ \bibnamefont {Loh}},\ }\bibfield  {title} {\enquote
  {\bibinfo {title} {Giant second-harmonic generation in ferroelectric
  {NbOI}$_2$},}\ }\href {\doibase 10.1038/s41566-022-01021-y} {\bibfield
  {journal} {\bibinfo  {journal} {Nat.\ Photonics}\ }\textbf {\bibinfo {volume}
  {16}},\ \bibinfo {pages} {644--650} (\bibinfo {year} {2022})}\BibitemShut
  {NoStop}%
\bibitem [{\citenamefont {Zograf}\ \emph {et~al.}(2024)\citenamefont {Zograf},
  \citenamefont {Polyakov}, \citenamefont {Bancerek}, \citenamefont
  {Antosiewicz}, \citenamefont {K\"u\c{c}\"uk\"oz},\ and\ \citenamefont
  {Shegai}}]{ZAY24}%
  \BibitemOpen
  \bibfield  {author} {\bibinfo {author} {\bibfnamefont {G.}~\bibnamefont
  {Zograf}}, \bibinfo {author} {\bibfnamefont {A.~Y.}\ \bibnamefont
  {Polyakov}}, \bibinfo {author} {\bibfnamefont {M.}~\bibnamefont {Bancerek}},
  \bibinfo {author} {\bibfnamefont {T.~J.}\ \bibnamefont {Antosiewicz}},
  \bibinfo {author} {\bibfnamefont {B.}~\bibnamefont {K\"u\c{c}\"uk\"oz}}, \
  and\ \bibinfo {author} {\bibfnamefont {T.~O.}\ \bibnamefont {Shegai}},\
  }\bibfield  {title} {\enquote {\bibinfo {title} {Combining ultrahigh index
  with exceptional nonlinearity in resonant transition metal dichalcogenide
  nanodisks},}\ }\href {\doibase 10.1038/s41566-024-01444-9} {\bibfield
  {journal} {\bibinfo  {journal} {Nat.\ Photonics}\ }\textbf {\bibinfo {volume}
  {18}},\ \bibinfo {pages} {751--757} (\bibinfo {year} {2024})}\BibitemShut
  {NoStop}%
\bibitem [{\citenamefont {Bernhardt}\ \emph {et~al.}(2020)\citenamefont
  {Bernhardt}, \citenamefont {Koshelev}, \citenamefont {White}, \citenamefont
  {Meng}, \citenamefont {Fr{\"o}ch}, \citenamefont {Kim}, \citenamefont {Tran},
  \citenamefont {Choi}, \citenamefont {Kivshar},\ and\ \citenamefont
  {Solntsev}}]{BKW20}%
  \BibitemOpen
  \bibfield  {author} {\bibinfo {author} {\bibfnamefont {N.}~\bibnamefont
  {Bernhardt}}, \bibinfo {author} {\bibfnamefont {K.}~\bibnamefont {Koshelev}},
  \bibinfo {author} {\bibfnamefont {S.~J.~U.}\ \bibnamefont {White}}, \bibinfo
  {author} {\bibfnamefont {K.~Wong~Choon}\ \bibnamefont {Meng}}, \bibinfo
  {author} {\bibfnamefont {J.~E.}\ \bibnamefont {Fr{\"o}ch}}, \bibinfo {author}
  {\bibfnamefont {S.}~\bibnamefont {Kim}}, \bibinfo {author} {\bibfnamefont
  {T.~T.}\ \bibnamefont {Tran}}, \bibinfo {author} {\bibfnamefont {D.-Y.}\
  \bibnamefont {Choi}}, \bibinfo {author} {\bibfnamefont {Y.}~\bibnamefont
  {Kivshar}}, \ and\ \bibinfo {author} {\bibfnamefont {A.~S.}\ \bibnamefont
  {Solntsev}},\ }\bibfield  {title} {\enquote {\bibinfo {title} {Quasi-{BIC}
  resonant enhancement of second-harmonic generation in {WS}$_2$ monolayers},}\
  }\href {\doibase 10.1021/acs.nanolett.0c01603} {\bibfield  {journal}
  {\bibinfo  {journal} {Nano\ Lett.}\ }\textbf {\bibinfo {volume} {20}},\
  \bibinfo {pages} {5309--5314} (\bibinfo {year} {2020})}\BibitemShut {NoStop}%
\bibitem [{\citenamefont {Pedersen}\ \emph {et~al.}(1999)\citenamefont
  {Pedersen}, \citenamefont {Pedersen},\ and\ \citenamefont
  {Kristensen}}]{PPK99}%
  \BibitemOpen
  \bibfield  {author} {\bibinfo {author} {\bibfnamefont {T.~G.}\ \bibnamefont
  {Pedersen}}, \bibinfo {author} {\bibfnamefont {K.}~\bibnamefont {Pedersen}},
  \ and\ \bibinfo {author} {\bibfnamefont {T.~B.}\ \bibnamefont {Kristensen}},\
  }\bibfield  {title} {\enquote {\bibinfo {title} {Optical second-harmonic
  generation from {Ag} quantum wells on {Si}(111)7$\times$7: {Experiment} and
  theory},}\ }\href {\doibase 10.1103/PhysRevB.60.R13997} {\bibfield  {journal}
  {\bibinfo  {journal} {Phys.\ Rev.\ B}\ }\textbf {\bibinfo {volume} {60}},\
  \bibinfo {pages} {R13997--R14000} (\bibinfo {year} {1999})}\BibitemShut
  {NoStop}%
\bibitem [{\citenamefont {Hirayama}\ \emph {et~al.}(2001)\citenamefont
  {Hirayama}, \citenamefont {Kawata},\ and\ \citenamefont
  {Takayanagi}}]{HKT01}%
  \BibitemOpen
  \bibfield  {author} {\bibinfo {author} {\bibfnamefont {H.}~\bibnamefont
  {Hirayama}}, \bibinfo {author} {\bibfnamefont {T.}~\bibnamefont {Kawata}}, \
  and\ \bibinfo {author} {\bibfnamefont {K.}~\bibnamefont {Takayanagi}},\
  }\bibfield  {title} {\enquote {\bibinfo {title} {Oscillation of the optical
  second-harmonic generation intensity during {Ag} thin film growth on a
  {Si}(111) 7$\times$7 surface},}\ }\href {\doibase 10.1103/PhysRevB.64.195415}
  {\bibfield  {journal} {\bibinfo  {journal} {Phys.\ Rev.\ B}\ }\textbf
  {\bibinfo {volume} {64}},\ \bibinfo {pages} {195415} (\bibinfo {year}
  {2001})}\BibitemShut {NoStop}%
\bibitem [{\citenamefont {Pedersen}\ \emph {et~al.}(2006)\citenamefont
  {Pedersen}, \citenamefont {Pedersen},\ and\ \citenamefont {Morgen}}]{PPM06}%
  \BibitemOpen
  \bibfield  {author} {\bibinfo {author} {\bibfnamefont {K.}~\bibnamefont
  {Pedersen}}, \bibinfo {author} {\bibfnamefont {T.~G.}\ \bibnamefont
  {Pedersen}}, \ and\ \bibinfo {author} {\bibfnamefont {P.}~\bibnamefont
  {Morgen}},\ }\bibfield  {title} {\enquote {\bibinfo {title} {Surface and
  interface resonances in second harmonic generation from metallic quantum
  wells on si(111)},}\ }\href {\doibase 10.1103/PhysRevB.73.125440} {\bibfield
  {journal} {\bibinfo  {journal} {Phys.\ Rev.\ B}\ }\textbf {\bibinfo {volume}
  {73}},\ \bibinfo {pages} {125440} (\bibinfo {year} {2006})}\BibitemShut
  {NoStop}%
\bibitem [{\citenamefont {{Rodr\'{\i}guez Echarri}}\ \emph
  {et~al.}(2021)\citenamefont {{Rodr\'{\i}guez Echarri}}, \citenamefont {Cox},
  \citenamefont {Iyikanat},\ and\ \citenamefont {{Garc\'{\i}a de
  Abajo}}}]{paper382}%
  \BibitemOpen
  \bibfield  {author} {\bibinfo {author} {\bibfnamefont {A.}~\bibnamefont
  {{Rodr\'{\i}guez Echarri}}}, \bibinfo {author} {\bibfnamefont {J.~D.}\
  \bibnamefont {Cox}}, \bibinfo {author} {\bibfnamefont {F.}~\bibnamefont
  {Iyikanat}}, \ and\ \bibinfo {author} {\bibfnamefont {F.~J.}\ \bibnamefont
  {{Garc\'{\i}a de Abajo}}},\ }\bibfield  {title} {\enquote {\bibinfo {title}
  {Nonlinear plasmonic response in atomically thin metal films},}\ }\href
  {\doibase 10.1515/nanoph-2021-0422} {\bibfield  {journal} {\bibinfo
  {journal} {Nanophotonics}\ }\textbf {\bibinfo {volume} {10}},\ \bibinfo
  {pages} {4149--4159} (\bibinfo {year} {2021})}\BibitemShut {NoStop}%
\bibitem [{\citenamefont {Pan}\ \emph {et~al.}(2024)\citenamefont {Pan},
  \citenamefont {Tong}, \citenamefont {Qian}, \citenamefont {Krasavin},
  \citenamefont {Li}, \citenamefont {Zhu}, \citenamefont {Zhang}, \citenamefont
  {Cui}, \citenamefont {Li}, \citenamefont {Wu}, \citenamefont {Liu},
  \citenamefont {Li}, \citenamefont {Guo}, \citenamefont {Zayats},
  \citenamefont {Tong},\ and\ \citenamefont {Wang}}]{PTQ24}%
  \BibitemOpen
  \bibfield  {author} {\bibinfo {author} {\bibfnamefont {C.}~\bibnamefont
  {Pan}}, \bibinfo {author} {\bibfnamefont {Y.}~\bibnamefont {Tong}}, \bibinfo
  {author} {\bibfnamefont {H.}~\bibnamefont {Qian}}, \bibinfo {author}
  {\bibfnamefont {A.~V.}\ \bibnamefont {Krasavin}}, \bibinfo {author}
  {\bibfnamefont {J.}~\bibnamefont {Li}}, \bibinfo {author} {\bibfnamefont
  {J.}~\bibnamefont {Zhu}}, \bibinfo {author} {\bibfnamefont {Y.}~\bibnamefont
  {Zhang}}, \bibinfo {author} {\bibfnamefont {B.}~\bibnamefont {Cui}}, \bibinfo
  {author} {\bibfnamefont {Z.}~\bibnamefont {Li}}, \bibinfo {author}
  {\bibfnamefont {C.}~\bibnamefont {Wu}}, \bibinfo {author} {\bibfnamefont
  {L.}~\bibnamefont {Liu}}, \bibinfo {author} {\bibfnamefont {L.}~\bibnamefont
  {Li}}, \bibinfo {author} {\bibfnamefont {X.}~\bibnamefont {Guo}}, \bibinfo
  {author} {\bibfnamefont {A.~V.}\ \bibnamefont {Zayats}}, \bibinfo {author}
  {\bibfnamefont {L.}~\bibnamefont {Tong}}, \ and\ \bibinfo {author}
  {\bibfnamefont {P.}~\bibnamefont {Wang}},\ }\bibfield  {title} {\enquote
  {\bibinfo {title} {Large area single crystal gold of single nanometer
  thickness for nanophotonics},}\ }\href {\doibase 10.1038/s41467-024-47133-7}
  {\bibfield  {journal} {\bibinfo  {journal} {Nat.\ Commun.}\ }\textbf
  {\bibinfo {volume} {15}},\ \bibinfo {pages} {2840} (\bibinfo {year}
  {2024})}\BibitemShut {NoStop}%
\bibitem [{\citenamefont {Jenke}\ \emph {et~al.}(2026)\citenamefont {Jenke},
  \citenamefont {Abdullah}, \citenamefont {Weber}, \citenamefont
  {{Rodr\'{\i}guez Echarri}}, \citenamefont {Iyikanat}, \citenamefont
  {Mkhitaryan}, \citenamefont {Schiller}, \citenamefont {Ortega}, \citenamefont
  {Walther}, \citenamefont {{Garc\'{\i}a de Abajo}},\ and\ \citenamefont
  {Rozema}}]{paper465}%
  \BibitemOpen
  \bibfield  {author} {\bibinfo {author} {\bibfnamefont {P.~K.}\ \bibnamefont
  {Jenke}}, \bibinfo {author} {\bibfnamefont {S.}~\bibnamefont {Abdullah}},
  \bibinfo {author} {\bibfnamefont {A.~P.}\ \bibnamefont {Weber}}, \bibinfo
  {author} {\bibfnamefont {A.}~\bibnamefont {{Rodr\'{\i}guez Echarri}}},
  \bibinfo {author} {\bibfnamefont {F.}~\bibnamefont {Iyikanat}}, \bibinfo
  {author} {\bibfnamefont {V.}~\bibnamefont {Mkhitaryan}}, \bibinfo {author}
  {\bibfnamefont {F.}~\bibnamefont {Schiller}}, \bibinfo {author}
  {\bibfnamefont {J.~E.}\ \bibnamefont {Ortega}}, \bibinfo {author}
  {\bibfnamefont {P.}~\bibnamefont {Walther}}, \bibinfo {author} {\bibfnamefont
  {F.~J.}\ \bibnamefont {{Garc\'{\i}a de Abajo}}}, \ and\ \bibinfo {author}
  {\bibfnamefont {L.~A.}\ \bibnamefont {Rozema}},\ }\bibfield  {title}
  {\enquote {\bibinfo {title} {Few-atom-thick silver films for enhanced
  nanoscale nonlinear optics},}\ }\href {\doibase 10.1038/s41467-026-74804-4}
  {\bibfield  {journal} {\bibinfo  {journal} {Nat.\ Commun.}\ }\textbf
  {\bibinfo {volume} {17}},\ \bibinfo {pages} {8214} (\bibinfo {year}
  {2026})}\BibitemShut {NoStop}%
\bibitem [{\citenamefont {{Abd El-Fattah}}\ \emph {et~al.}(2019)\citenamefont
  {{Abd El-Fattah}}, \citenamefont {Mkhitaryan}, \citenamefont {Brede},
  \citenamefont {Fern\'andez}, \citenamefont {Li}, \citenamefont {Guo},
  \citenamefont {Ghosh}, \citenamefont {{Rodr\'{\i}guez Echarri}},
  \citenamefont {Naveh}, \citenamefont {Xia}, \citenamefont {Ortega},\ and\
  \citenamefont {{Garc\'{\i}a de Abajo}}}]{paper335}%
  \BibitemOpen
  \bibfield  {author} {\bibinfo {author} {\bibfnamefont {Z.~M.}\ \bibnamefont
  {{Abd El-Fattah}}}, \bibinfo {author} {\bibfnamefont {V.}~\bibnamefont
  {Mkhitaryan}}, \bibinfo {author} {\bibfnamefont {J.}~\bibnamefont {Brede}},
  \bibinfo {author} {\bibfnamefont {L.}~\bibnamefont {Fern\'andez}}, \bibinfo
  {author} {\bibfnamefont {C.}~\bibnamefont {Li}}, \bibinfo {author}
  {\bibfnamefont {Q.}~\bibnamefont {Guo}}, \bibinfo {author} {\bibfnamefont
  {A.}~\bibnamefont {Ghosh}}, \bibinfo {author} {\bibfnamefont
  {A.}~\bibnamefont {{Rodr\'{\i}guez Echarri}}}, \bibinfo {author}
  {\bibfnamefont {D.}~\bibnamefont {Naveh}}, \bibinfo {author} {\bibfnamefont
  {F.}~\bibnamefont {Xia}}, \bibinfo {author} {\bibfnamefont {J.~E.}\
  \bibnamefont {Ortega}}, \ and\ \bibinfo {author} {\bibfnamefont {F.~J.}\
  \bibnamefont {{Garc\'{\i}a de Abajo}}},\ }\bibfield  {title} {\enquote
  {\bibinfo {title} {Plasmonics in atomically thin crystalline silver films},}\
  }\href {\doibase 10.1021/acsnano.9b01651} {\bibfield  {journal} {\bibinfo
  {journal} {ACS\ Nano}\ }\textbf {\bibinfo {volume} {13}},\ \bibinfo {pages}
  {7771--7779} (\bibinfo {year} {2019})}\BibitemShut {NoStop}%
\bibitem [{\citenamefont {Mkhitaryan}\ \emph {et~al.}(2024)\citenamefont
  {Mkhitaryan}, \citenamefont {Weber}, \citenamefont {Abdullah}, \citenamefont
  {Fern\'andez}, \citenamefont {{Abd El-Fattah}}, \citenamefont
  {Piquero-Zulaica}, \citenamefont {Agarwal}, \citenamefont {{Garc\'{\i}a
  D\'{\i}ez}}, \citenamefont {Schiller}, \citenamefont {Ortega},\ and\
  \citenamefont {{Garc\'{\i}a de Abajo}}}]{paper427}%
  \BibitemOpen
  \bibfield  {author} {\bibinfo {author} {\bibfnamefont {V.}~\bibnamefont
  {Mkhitaryan}}, \bibinfo {author} {\bibfnamefont {A.~P.}\ \bibnamefont
  {Weber}}, \bibinfo {author} {\bibfnamefont {S.}\ \bibnamefont {Abdullah}},
  \bibinfo {author} {\bibfnamefont {L.}~\bibnamefont {Fern\'andez}}, \bibinfo
  {author} {\bibfnamefont {Z.~M.}\ \bibnamefont {{Abd El-Fattah}}}, \bibinfo
  {author} {\bibfnamefont {I.}~\bibnamefont {Piquero-Zulaica}}, \bibinfo
  {author} {\bibfnamefont {H.}~\bibnamefont {Agarwal}}, \bibinfo {author}
  {\bibfnamefont {K.}~\bibnamefont {{Garc\'{\i}a D\'{\i}ez}}}, \bibinfo
  {author} {\bibfnamefont {F.}~\bibnamefont {Schiller}}, \bibinfo {author}
  {\bibfnamefont {J.~E.}\ \bibnamefont {Ortega}}, \ and\ \bibinfo {author}
  {\bibfnamefont {F.~J.}\ \bibnamefont {{Garc\'{\i}a de Abajo}}},\ }\bibfield
  {title} {\enquote {\bibinfo {title} {Ultraconfined plasmons in atomically
  thin crystalline silver nanostructures},}\ }\href {\doibase
  10.1002/adma.202302520} {\bibfield  {journal} {\bibinfo  {journal} {Adv.\
  Mater.}\ }\textbf {\bibinfo {volume} {36}},\ \bibinfo {pages} {2302520}
  (\bibinfo {year} {2024})}\BibitemShut {NoStop}%
\bibitem [{\citenamefont {Stockman}(2011)}]{S11}%
  \BibitemOpen
  \bibfield  {author} {\bibinfo {author} {\bibfnamefont {M.~I.}\ \bibnamefont
  {Stockman}},\ }\bibfield  {title} {\enquote {\bibinfo {title}
  {Nanoplasmonics: the physics {behind} the applications},}\ }\href {\doibase
  10.1063/1.3554315} {\bibfield  {journal} {\bibinfo  {journal} {Phys.\ Today}\
  }\textbf {\bibinfo {volume} {64}},\ \bibinfo {pages} {39--44} (\bibinfo
  {year} {2011})}\BibitemShut {NoStop}%
\bibitem [{\citenamefont {Cesca}\ \emph {et~al.}(2010)\citenamefont {Cesca},
  \citenamefont {Pellegrini}, \citenamefont {Bello}, \citenamefont {Scian},
  \citenamefont {Mazzoldi}, \citenamefont {Calvelli}, \citenamefont
  {Battaglin},\ and\ \citenamefont {Mattei}}]{CPB10}%
  \BibitemOpen
  \bibfield  {author} {\bibinfo {author} {\bibfnamefont {T.}~\bibnamefont
  {Cesca}}, \bibinfo {author} {\bibfnamefont {G.}~\bibnamefont {Pellegrini}},
  \bibinfo {author} {\bibfnamefont {V.}~\bibnamefont {Bello}}, \bibinfo
  {author} {\bibfnamefont {C.}~\bibnamefont {Scian}}, \bibinfo {author}
  {\bibfnamefont {P.}~\bibnamefont {Mazzoldi}}, \bibinfo {author}
  {\bibfnamefont {P.}~\bibnamefont {Calvelli}}, \bibinfo {author}
  {\bibfnamefont {G.}~\bibnamefont {Battaglin}}, \ and\ \bibinfo {author}
  {\bibfnamefont {G.}~\bibnamefont {Mattei}},\ }\bibfield  {title} {\enquote
  {\bibinfo {title} {Nonlinear optical properties of Au--Ag nanoplanets made
  by ion beam processing of bimetallic nanoclusters in silica},}\ }\href
  {\doibase 10.1016/j.nimb.2010.05.095} {\bibfield  {journal} {\bibinfo
  {journal} {Nucl.\ Instrum.\ Methods\ Phys.\ Res.\ B}\ }\textbf {\bibinfo
  {volume} {268}},\ \bibinfo {pages} {3227--3230} (\bibinfo {year}
  {2010})}\BibitemShut {NoStop}%
\bibitem [{\citenamefont {Leopold}\ \emph {et~al.}(2013)\citenamefont
  {Leopold}, \citenamefont {Chi\c{s}}, \citenamefont {Mircescu}, \citenamefont
  {Mari\c{s}ca}, \citenamefont {Buja}, \citenamefont {Leopold}, \citenamefont
  {Socaciu}, \citenamefont {Braicu}, \citenamefont {Irimie},\ and\
  \citenamefont {Berindan-Neagoe}}]{LCM13}%
  \BibitemOpen
  \bibfield  {author} {\bibinfo {author} {\bibfnamefont {N.}~\bibnamefont
  {Leopold}}, \bibinfo {author} {\bibfnamefont {V.}~\bibnamefont {Chi\c{s}}},
  \bibinfo {author} {\bibfnamefont {N.~E.}\ \bibnamefont {Mircescu}}, \bibinfo
  {author} {\bibfnamefont {O.~T.}\ \bibnamefont {Mari\c{s}ca}}, \bibinfo
  {author} {\bibfnamefont {O.~M.}\ \bibnamefont {Buja}}, \bibinfo {author}
  {\bibfnamefont {L.~F.}\ \bibnamefont {Leopold}}, \bibinfo {author}
  {\bibfnamefont {C.}~\bibnamefont {Socaciu}}, \bibinfo {author} {\bibfnamefont
  {C.}~\bibnamefont {Braicu}}, \bibinfo {author} {\bibfnamefont
  {A.}~\bibnamefont {Irimie}}, \ and\ \bibinfo {author} {\bibfnamefont
  {I.}~\bibnamefont {Berindan-Neagoe}},\ }\bibfield  {title} {\enquote
  {\bibinfo {title} {One step synthesis of {SERS} active colloidal gold
  nanoparticles by reduction with polyethylene glycol},}\ }\href {\doibase
  10.1016/j.colsurfa.2013.05.075} {\bibfield  {journal} {\bibinfo  {journal}
  {Colloids\ Surf.\ A}\ }\textbf {\bibinfo {volume} {436}},\ \bibinfo {pages}
  {133--138} (\bibinfo {year} {2013})}\BibitemShut {NoStop}%
\bibitem [{\citenamefont {Ahmed}\ \emph {et~al.}(2017)\citenamefont {Ahmed},
  \citenamefont {Kim}, \citenamefont {Tran}, \citenamefont {Suzuki},
  \citenamefont {Neethirajan}, \citenamefont {Lee},\ and\ \citenamefont
  {Park}}]{AKT17}%
  \BibitemOpen
  \bibfield  {author} {\bibinfo {author} {\bibfnamefont {S.~R.}\ \bibnamefont
  {Ahmed}}, \bibinfo {author} {\bibfnamefont {J.}~\bibnamefont {Kim}}, \bibinfo
  {author} {\bibfnamefont {V.~T.}\ \bibnamefont {Tran}}, \bibinfo {author}
  {\bibfnamefont {T.}~\bibnamefont {Suzuki}}, \bibinfo {author} {\bibfnamefont
  {S.}~\bibnamefont {Neethirajan}}, \bibinfo {author} {\bibfnamefont
  {J.}~\bibnamefont {Lee}}, \ and\ \bibinfo {author} {\bibfnamefont {E.~Y.}\
  \bibnamefont {Park}},\ }\bibfield  {title} {\enquote {\bibinfo {title} {In
  situ self-assembly of gold nanoparticles on hydrophilic and hydrophobic
  substrates for influenza virus-sensing platform},}\ }\href {\doibase
  10.1038/srep44495} {\bibfield  {journal} {\bibinfo  {journal} {Sci.\ Rep.}\
  }\textbf {\bibinfo {volume} {7}},\ \bibinfo {pages} {44495} (\bibinfo {year}
  {2017})}\BibitemShut {NoStop}%
\bibitem [{\citenamefont {Chiang}\ \emph {et~al.}(2019)\citenamefont {Chiang},
  \citenamefont {Wang}, \citenamefont {Zhang},\ and\ \citenamefont
  {Levon}}]{CWZ19}%
  \BibitemOpen
  \bibfield  {author} {\bibinfo {author} {\bibfnamefont {H.-C.}\ \bibnamefont
  {Chiang}}, \bibinfo {author} {\bibfnamefont {Y.}~\bibnamefont {Wang}},
  \bibinfo {author} {\bibfnamefont {Q.}~\bibnamefont {Zhang}}, \ and\ \bibinfo
  {author} {\bibfnamefont {K.}~\bibnamefont {Levon}},\ }\bibfield  {title}
  {\enquote {\bibinfo {title} {Optimization of the electrodeposition of gold
  nanoparticles for the application of highly sensitive, label-free
  biosensor},}\ }\href {\doibase 10.3390/bios9020050} {\bibfield  {journal}
  {\bibinfo  {journal} {Biosensors}\ }\textbf {\bibinfo {volume} {9}},\
  \bibinfo {pages} {50} (\bibinfo {year} {2019})}\BibitemShut {NoStop}%
\bibitem [{\citenamefont {Parkhomenko}\ \emph {et~al.}(2015)\citenamefont
  {Parkhomenko}, \citenamefont {Trubin}, \citenamefont {Turgambaeva},\ and\
  \citenamefont {Igumenov}}]{PTT15}%
  \BibitemOpen
  \bibfield  {author} {\bibinfo {author} {\bibfnamefont {R.~G.}\ \bibnamefont
  {Parkhomenko}}, \bibinfo {author} {\bibfnamefont {S.~V.}\ \bibnamefont
  {Trubin}}, \bibinfo {author} {\bibfnamefont {A.~E.}\ \bibnamefont
  {Turgambaeva}}, \ and\ \bibinfo {author} {\bibfnamefont {I.~K.}\ \bibnamefont
  {Igumenov}},\ }\bibfield  {title} {\enquote {\bibinfo {title} {Deposition of
  pure gold thin films from organometallic precursors},}\ }\href {\doibase
  10.1016/j.jcrysgro.2014.09.034} {\bibfield  {journal} {\bibinfo  {journal}
  {J.\ Cryst.\ Growth}\ }\textbf {\bibinfo {volume} {414}},\ \bibinfo {pages}
  {143--150} (\bibinfo {year} {2015})}\BibitemShut {NoStop}%
\bibitem [{\citenamefont {McPeak}\ \emph {et~al.}(2015)\citenamefont {McPeak},
  \citenamefont {Jayanti}, \citenamefont {Kress}, \citenamefont {Meyer},
  \citenamefont {Iotti}, \citenamefont {Rossinelli},\ and\ \citenamefont
  {Norris}}]{MJK15}%
  \BibitemOpen
  \bibfield  {author} {\bibinfo {author} {\bibfnamefont {K.~M.}\ \bibnamefont
  {McPeak}}, \bibinfo {author} {\bibfnamefont {S.~V.}\ \bibnamefont {Jayanti}},
  \bibinfo {author} {\bibfnamefont {S.~J.~P.}\ \bibnamefont {Kress}}, \bibinfo
  {author} {\bibfnamefont {S.}~\bibnamefont {Meyer}}, \bibinfo {author}
  {\bibfnamefont {S.}~\bibnamefont {Iotti}}, \bibinfo {author} {\bibfnamefont
  {A.}~\bibnamefont {Rossinelli}}, \ and\ \bibinfo {author} {\bibfnamefont
  {D.~J.}\ \bibnamefont {Norris}},\ }\bibfield  {title} {\enquote {\bibinfo
  {title} {Plasmonic films can easily be better: rules and recipes},}\ }\href
  {\doibase 10.1021/ph5004237} {\bibfield  {journal} {\bibinfo  {journal} {ACS\
  Photonics}\ }\textbf {\bibinfo {volume} {2}},\ \bibinfo {pages} {326--333}
  (\bibinfo {year} {2015})}\BibitemShut {NoStop}%
\bibitem [{\citenamefont {Sun}\ \emph {et~al.}(2007)\citenamefont {Sun},
  \citenamefont {Hong}, \citenamefont {Hou}, \citenamefont {Fan},\ and\
  \citenamefont {Shao}}]{SHH07}%
  \BibitemOpen
  \bibfield  {author} {\bibinfo {author} {\bibfnamefont {X.}~\bibnamefont
  {Sun}}, \bibinfo {author} {\bibfnamefont {R.}~\bibnamefont {Hong}},
  \bibinfo {author} {\bibfnamefont {H.}~\bibnamefont {Hou}}, \bibinfo {author}
  {\bibfnamefont {Z.}~\bibnamefont {Fan}}, \ and\ \bibinfo {author}
  {\bibfnamefont {J.}~\bibnamefont {Shao}},\ }\bibfield  {title} {\enquote
  {\bibinfo {title} {Thickness dependence of structure and optical properties
  of silver films deposited by magnetron sputtering},}\ }\href {\doibase
  10.1016/j.tsf.2007.02.017} {\bibfield  {journal} {\bibinfo  {journal} {Thin
  Solid Films}\ }\textbf {\bibinfo {volume} {515}},\ \bibinfo {pages}
  {6962--6966} (\bibinfo {year} {2007})}\BibitemShut {NoStop}%
\bibitem [{\citenamefont {Kiani}\ and\ \citenamefont {Tagliabue}(2022)}]{KT22}%
  \BibitemOpen
  \bibfield  {author} {\bibinfo {author} {\bibfnamefont {F.}~\bibnamefont
  {Kiani}}\ and\ \bibinfo {author} {\bibfnamefont {G.}~\bibnamefont
  {Tagliabue}},\ }\bibfield  {title} {\enquote {\bibinfo {title} {High aspect
  ratio {Au} microflakes {via} gap-assisted synthesis},}\ }\href {\doibase
  10.1021/acs.chemmater.1c03908} {\bibfield  {journal} {\bibinfo  {journal}
  {Chem.\ Mater.}\ }\textbf {\bibinfo {volume} {34}},\ \bibinfo {pages}
  {1278--1288} (\bibinfo {year} {2022})}\BibitemShut {NoStop}%
\bibitem [{\citenamefont {Speer}\ \emph {et~al.}(2006)\citenamefont {Speer},
  \citenamefont {Tang}, \citenamefont {Miller},\ and\ \citenamefont
  {Chiang}}]{STM06}%
  \BibitemOpen
  \bibfield  {author} {\bibinfo {author} {\bibfnamefont {N.~J.}\ \bibnamefont
  {Speer}}, \bibinfo {author} {\bibfnamefont {S.-J.}\ \bibnamefont {Tang}},
  \bibinfo {author} {\bibfnamefont {T.}~\bibnamefont {Miller}}, \ and\ \bibinfo
  {author} {\bibfnamefont {T.-C.}\ \bibnamefont {Chiang}},\ }\bibfield  {title}
  {\enquote {\bibinfo {title} {Coherent electronic fringe structure in
  incommensurate silver-silicon quantum wells},}\ }\href {\doibase
  10.1126/science.1132941} {\bibfield  {journal} {\bibinfo  {journal}
  {Science}\ }\textbf {\bibinfo {volume} {314}},\ \bibinfo {pages} {804--806}
  (\bibinfo {year} {2006})}\BibitemShut {NoStop}%
\bibitem [{\citenamefont {Butet}\ \emph {et~al.}(2015)\citenamefont {Butet},
  \citenamefont {Brevet},\ and\ \citenamefont {Martin}}]{BBM15}%
  \BibitemOpen
  \bibfield  {author} {\bibinfo {author} {\bibfnamefont {J.}~\bibnamefont
  {Butet}}, \bibinfo {author} {\bibfnamefont {P.-F.}\ \bibnamefont {Brevet}}, \
  and\ \bibinfo {author} {\bibfnamefont {O.~J.~F.}\ \bibnamefont {Martin}},\
  }\bibfield  {title} {\enquote {\bibinfo {title} {Optical second harmonic
  generation in plasmonic nanostructures: from fundamental principles to
  advanced applications},}\ }\href {\doibase 10.1021/acsnano.5b04373}
  {\bibfield  {journal} {\bibinfo  {journal} {ACS\ Nano}\ }\textbf {\bibinfo
  {volume} {9}},\ \bibinfo {pages} {10545--10562} (\bibinfo {year}
  {2015})}\BibitemShut {NoStop}%
\bibitem [{\citenamefont {Rossetti}\ \emph {et~al.}(2025)\citenamefont
  {Rossetti}, \citenamefont {Hu}, \citenamefont {Venanzi}, \citenamefont
  {Boussekso}, \citenamefont {{De Luca}}, \citenamefont {Deckert},
  \citenamefont {Giliberti}, \citenamefont {Pea}, \citenamefont {Sagnes},
  \citenamefont {Beaudoin}, \citenamefont {Biagioni}, \citenamefont {Ba\`u},
  \citenamefont {Maier}, \citenamefont {Tittl}, \citenamefont {Brida},
  \citenamefont {Colombelli}, \citenamefont {Ortolani},\ and\ \citenamefont
  {Cirac\`i}}]{RHV25}%
  \BibitemOpen
  \bibfield  {author} {\bibinfo {author} {\bibfnamefont {A.}~\bibnamefont
  {Rossetti}}, \bibinfo {author} {\bibfnamefont {H.}~\bibnamefont {Hu}},
  \bibinfo {author} {\bibfnamefont {T.}~\bibnamefont {Venanzi}}, \bibinfo
  {author} {\bibfnamefont {A.}~\bibnamefont {Boussekso}}, \bibinfo {author}
  {\bibfnamefont {F.}~\bibnamefont {{De Luca}}}, \bibinfo {author}
  {\bibfnamefont {T.}~\bibnamefont {Deckert}}, \bibinfo {author} {\bibfnamefont
  {V.}~\bibnamefont {Giliberti}}, \bibinfo {author} {\bibfnamefont
  {M.}~\bibnamefont {Pea}}, \bibinfo {author} {\bibfnamefont {I.}~\bibnamefont
  {Sagnes}}, \bibinfo {author} {\bibfnamefont {G.}~\bibnamefont {Beaudoin}},
  \bibinfo {author} {\bibfnamefont {P.}~\bibnamefont {Biagioni}}, \bibinfo
  {author} {\bibfnamefont {E.}~\bibnamefont {Ba\`u}}, \bibinfo {author}
  {\bibfnamefont {S.~A.}\ \bibnamefont {Maier}}, \bibinfo {author}
  {\bibfnamefont {A.}~\bibnamefont {Tittl}}, \bibinfo {author} {\bibfnamefont
  {D.}~\bibnamefont {Brida}}, \bibinfo {author} {\bibfnamefont
  {R.}~\bibnamefont {Colombelli}}, \bibinfo {author} {\bibfnamefont
  {M.}~\bibnamefont {Ortolani}}, \ and\ \bibinfo {author} {\bibfnamefont
  {C.}~\bibnamefont {Cirac\`i}},\ }\bibfield  {title} {\enquote {\bibinfo
  {title} {Control and enhancement of optical nonlinearities in plasmonic
  semiconductor nanostructures},}\ }\href {\doibase 10.1038/s41377-025-01783-4}
  {\bibfield  {journal} {\bibinfo  {journal} {Light\ Sci.\ Appl.}\ }\textbf
  {\bibinfo {volume} {14}},\ \bibinfo {pages} {192} (\bibinfo {year}
  {2025})}\BibitemShut {NoStop}%
\bibitem [{\citenamefont {Mesch}\ \emph {et~al.}(2016)\citenamefont {Mesch},
  \citenamefont {Metzger}, \citenamefont {Hentschel},\ and\ \citenamefont
  {Giessen}}]{MMH16}%
  \BibitemOpen
  \bibfield  {author} {\bibinfo {author} {\bibfnamefont {M.}~\bibnamefont
  {Mesch}}, \bibinfo {author} {\bibfnamefont {B.}~\bibnamefont {Metzger}},
  \bibinfo {author} {\bibfnamefont {M.}~\bibnamefont {Hentschel}}, \ and\
  \bibinfo {author} {\bibfnamefont {H.}~\bibnamefont {Giessen}},\ }\bibfield
  {title} {\enquote {\bibinfo {title} {Nonlinear plasmonic sensing},}\ }\href
  {\doibase 10.1021/acs.nanolett.6b00478} {\bibfield  {journal} {\bibinfo
  {journal} {Nano\ Lett.}\ }\textbf {\bibinfo {volume} {16}},\ \bibinfo {pages}
  {3155--3159} (\bibinfo {year} {2016})}\BibitemShut {NoStop}%
\bibitem [{\citenamefont {Schiller}\ \emph {et~al.}(2005)\citenamefont
  {Schiller}, \citenamefont {Cord\'on}, \citenamefont {Vyalikh}, \citenamefont
  {Rubio},\ and\ \citenamefont {Ortega}}]{SCV05}%
  \BibitemOpen
  \bibfield  {author} {\bibinfo {author} {\bibfnamefont {F.}~\bibnamefont
  {Schiller}}, \bibinfo {author} {\bibfnamefont {J.}~\bibnamefont {Cord\'on}},
  \bibinfo {author} {\bibfnamefont {D.}~\bibnamefont {Vyalikh}}, \bibinfo
  {author} {\bibfnamefont {A.}~\bibnamefont {Rubio}}, \ and\ \bibinfo {author}
  {\bibfnamefont {J.~E.}\ \bibnamefont {Ortega}},\ }\bibfield  {title}
  {\enquote {\bibinfo {title} {{Fermi} gap stabilization of an incommensurate
  two-dimensional superstructure},}\ }\href {\doibase
  10.1103/PhysRevLett.94.016103} {\bibfield  {journal} {\bibinfo  {journal}
  {Phys.\ Rev.\ Lett.}\ }\textbf {\bibinfo {volume} {94}},\ \bibinfo {pages}
  {016103} (\bibinfo {year} {2005})}\BibitemShut {NoStop}%
\bibitem [{\citenamefont {Fukumoto}\ \emph {et~al.}(2013)\citenamefont
  {Fukumoto}, \citenamefont {Miyazaki}, \citenamefont {Aoki}, \citenamefont
  {Nakatsuji},\ and\ \citenamefont {Hirayama}}]{FMA13}%
  \BibitemOpen
  \bibfield  {author} {\bibinfo {author} {\bibfnamefont {H.}~\bibnamefont
  {Fukumoto}}, \bibinfo {author} {\bibfnamefont {M.}~\bibnamefont {Miyazaki}},
  \bibinfo {author} {\bibfnamefont {Y.}~\bibnamefont {Aoki}}, \bibinfo {author}
  {\bibfnamefont {K.}~\bibnamefont {Nakatsuji}}, \ and\ \bibinfo {author}
  {\bibfnamefont {H.}~\bibnamefont {Hirayama}},\ }\bibfield  {title} {\enquote
  {\bibinfo {title} {Initial stage of {Ag} growth on
  {Bi/Ag}(111)$\sqrt{3}\times\sqrt{3}$ surfaces},}\ }\href {\doibase
  10.1016/j.susc.2013.01.013} {\bibfield  {journal} {\bibinfo  {journal}
  {Surf.\ Sci.}\ }\textbf {\bibinfo {volume} {611}},\ \bibinfo {pages} {49--53}
  (\bibinfo {year} {2013})}\BibitemShut {NoStop}%
\bibitem [{\citenamefont {Franta}\ \emph {et~al.}(2018)\citenamefont {Franta},
  \citenamefont {Franta}, \citenamefont {Voh{\'a}nka}, \citenamefont
  {{\v{C}}erm{\'a}k},\ and\ \citenamefont {Ohl{\'\i}dal}}]{FFV18}%
  \BibitemOpen
  \bibfield  {author} {\bibinfo {author} {\bibfnamefont {D.}~\bibnamefont
  {Franta}}, \bibinfo {author} {\bibfnamefont {P.}~\bibnamefont {Franta}},
  \bibinfo {author} {\bibfnamefont {J.}~\bibnamefont {Voh{\'a}nka}}, \bibinfo
  {author} {\bibfnamefont {M.}~\bibnamefont {{\v{C}}erm{\'a}k}}, \ and\
  \bibinfo {author} {\bibfnamefont {I.}~\bibnamefont {Ohl{\'\i}dal}},\
  }\bibfield  {title} {\enquote {\bibinfo {title} {Determination of thicknesses
  and temperatures of crystalline silicon wafers from optical measurements in
  the far infrared region},}\ }\href {\doibase 10.1063/1.5026195} {\bibfield
  {journal} {\bibinfo  {journal} {J.\ Appl.\ Phys.}\ }\textbf {\bibinfo
  {volume} {123}},\ \bibinfo {pages} {185707} (\bibinfo {year}
  {2018})}\BibitemShut {NoStop}%
\bibitem [{\citenamefont {Novotny}\ and\ \citenamefont {Hecht}(2006)}]{NH06}%
  \BibitemOpen
  \bibfield  {author} {\bibinfo {author} {\bibfnamefont {L.}~\bibnamefont
  {Novotny}}\ and\ \bibinfo {author} {\bibfnamefont {B.}~\bibnamefont
  {Hecht}},\ }\href@noop {} {\emph {\bibinfo {title} {Principles of
  {N}ano-{O}ptics}}}\ (\bibinfo  {publisher} {Cambridge University Press},\
  \bibinfo {address} {New York},\ \bibinfo {year} {2006})\BibitemShut {NoStop}%
\bibitem [{\citenamefont {Johnson}\ and\ \citenamefont
  {Christy}(1972)}]{JC1972}%
  \BibitemOpen
  \bibfield  {author} {\bibinfo {author} {\bibfnamefont {P.~B.}\ \bibnamefont
  {Johnson}}\ and\ \bibinfo {author} {\bibfnamefont {R.~W.}\ \bibnamefont
  {Christy}},\ }\bibfield  {title} {\enquote {\bibinfo {title} {Optical
  constants of the noble metals},}\ }\href {\doibase 10.1103/PhysRevB.6.4370}
  {\bibfield  {journal} {\bibinfo  {journal} {Phys.\ Rev.\ B}\ }\textbf
  {\bibinfo {volume} {6}},\ \bibinfo {pages} {4370--4379} (\bibinfo {year}
  {1972})}\BibitemShut {NoStop}%
\end{thebibliography}

%

\pagebreak \onecolumngrid \section*{SUPPLEMENTARY FIGURES}
\renewcommand{\thefigure}{S\arabic{figure}}
\setcounter{figure}{0}

\begin{figure*}[hpbt]
\centering\includegraphics[width=1.0\linewidth]{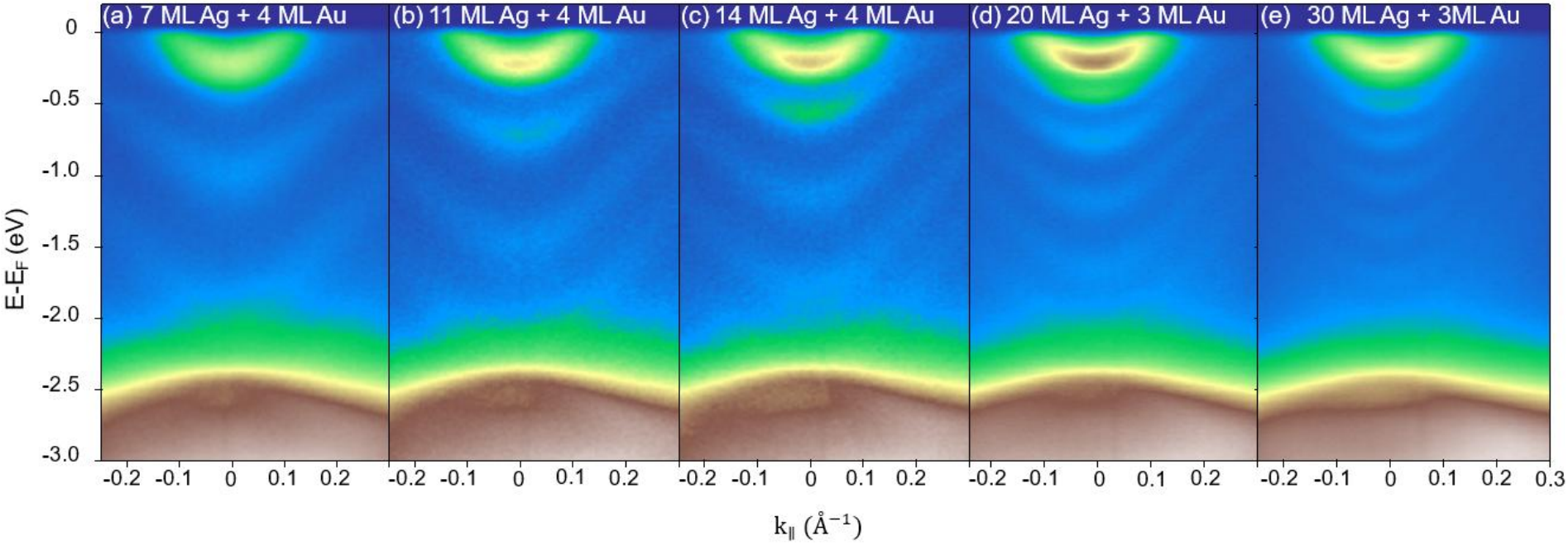}
\caption{\textbf{Angle-resolved photoemission spectroscopy (ARPES) of ultrathin Ag films with Au capping.} We plot measured ARPES intensity maps as a function of binding energy $E$ relative to the Fermi energy $\EF$ and in-plane wave vector $k_\parallel$ for atomically thin Ag films of different thicknesses: (a)~7~ML Ag + 4~ML Au, (b)~11~ML Ag + 4~ML Au, (c)~14~ML Ag + 4~ML Au, (d)~20~ML Ag + 3~ML Au, and (e)~30~ML Ag + 3~ML Au. The data reveal the evolution of the thickness-dependent electronic structure and quantum-well states.}
\label{FigS1}
\end{figure*}

\begin{figure*}[hpbt]
\centering\includegraphics[width=0.95\linewidth]{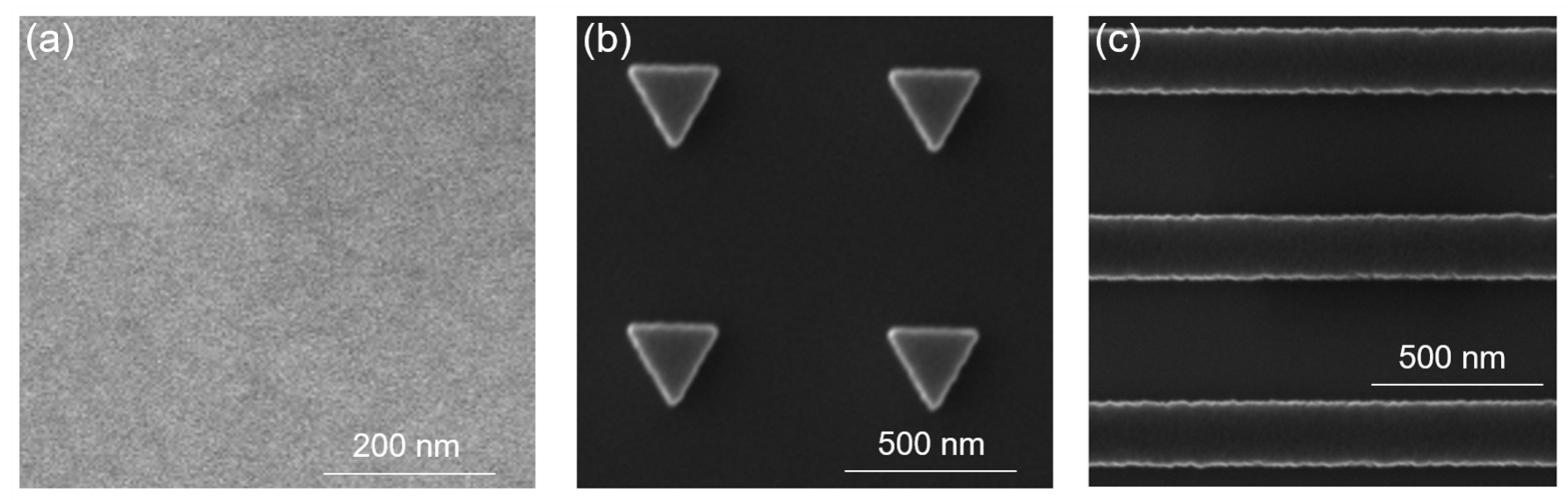}
\caption{\textbf{Scanning electron microscopy (SEM) images of a planar film and fabricated nanostructures.} (a)~Unpatterned 11~ML Ag film passivated with 4~MLs of Au, showing a smooth and uniform surface morphology. (b)~Periodic array of triangular nanostructures with a side length of $\sim$300~nm. (c)~Parallel nanoribbons with a width of $\sim$200~nm. The lattice period in (b) and (c) is approximately three times the characteristic lateral dimension of each nanostructure. All structures are fabricated from the same film. Scale bars: (a)~200~nm and (b,c)~500~nm.}
\label{FigS2}
\end{figure*}

\begin{figure*}[hpbt]
\centering\includegraphics[width=0.9\linewidth]{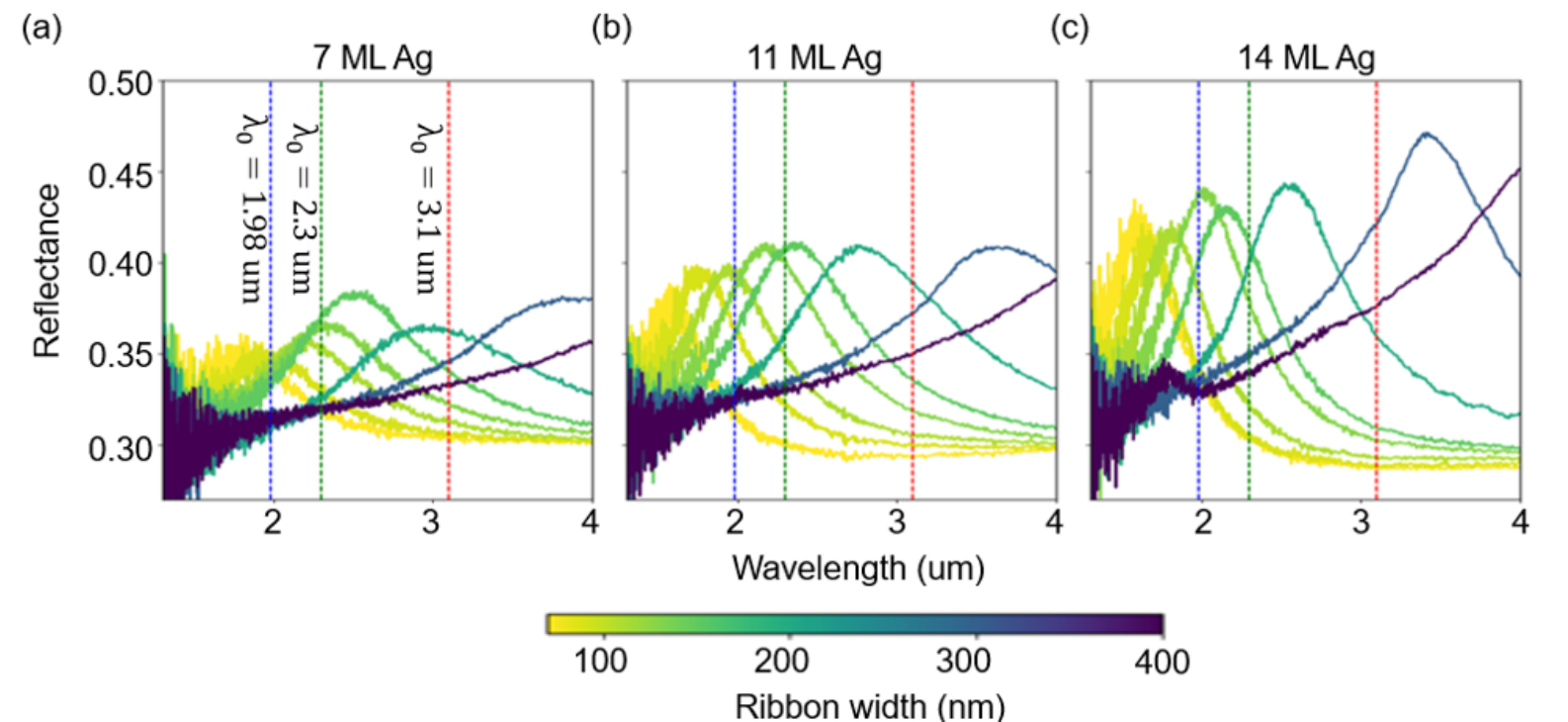}
\caption{\textbf{Optical characterization of silver nanoribbons.} We plot FTIR measurements of reflectance spectra for samples consisting of (a)~7~ML, (b)~11~ML, and (c)~14~ML Ag films, each of them passivated with 4~MLs of Au. Curves correspond to different ribbon widths ranging from 100~nm to 400~nm, as indicated by the color scale. Vertical dashed lines mark the fundamental excitation wavelengths $\lambda_0=1.98~\um$ (blue), $2.3~\um$ (green), and $3.1~\um$ (red) used in the SHG measurements.}
\label{FigS3}
\end{figure*}

\begin{figure}[hpbt]
\centering\includegraphics[width=0.7\linewidth]{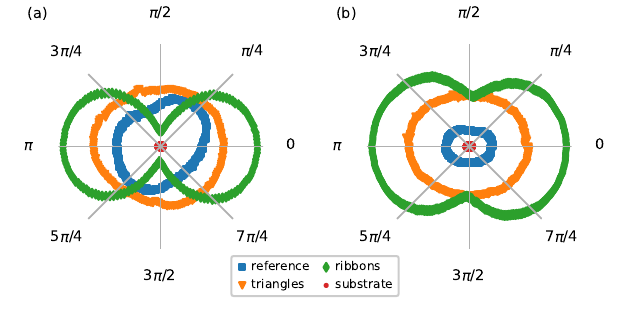}
\caption{\textbf{Polarization-dependent second-harmonic-generation (SHG) measurements for ribbons and reference structures.} (a)~Input polarization scan: the excitation polarization was rotated $360^\circ$ relative to the long axis of the ribbons, while the SHG count rate was integrated over all output polarizations. (b)~Output polarization scan: the excitation polarization was fixed to horizontal polarization, and the emitted SHG signal was analyzed by selectively detecting different polarization components. Data are shown for the planar metal reference film (blue squares), ribbons with 200~nm width (green diamonds), triangles with 200~nm side length (orange triangles), and the bare silicon substrate (red circles). Measurements were performed on a sample consisting of 7~MLs of Ag capped with 4~MLs of Au.}
\label{FigS4}
\end{figure}

\begin{figure}[hpbt]
\centering\includegraphics[width=0.9\linewidth]{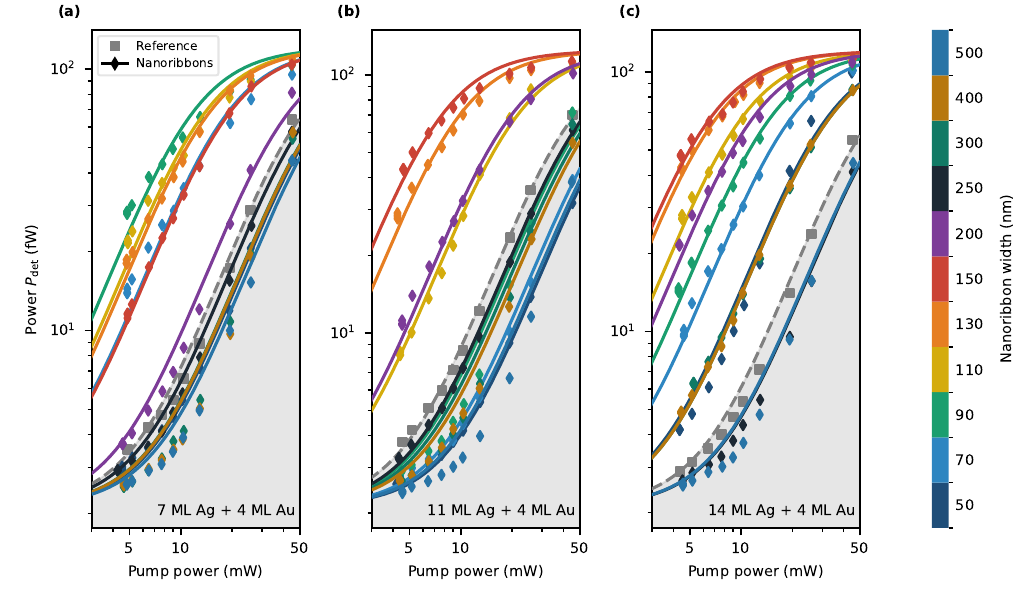}
\caption{\textbf{Power scaling of the measured second-harmonic signal at an excitation wavelength of $2.3~\um$.} We plot the measured SHG average power as a function of average pump power (symbols) for nanoribbon arrays as described in the main text. Different ribbon widths are considered, as encoded in the right color scale. We show results for three different film thicknesses: (a)~7~ML Ag, (b)~11~ML Ag, and (c)~14~ML Ag, all capped with 4~MLs of Au. Data for the unstructured film are shown for reference (grey shading). Curves show second-order power-law fits. The excitation average power and the SHG power are measured before and after the sample, respectively. Fresnel reflections are not taken into account.}
\label{FigS5}
\end{figure}

\begin{figure*}[hpbt]
\centering
\includegraphics[width=0.95\linewidth]{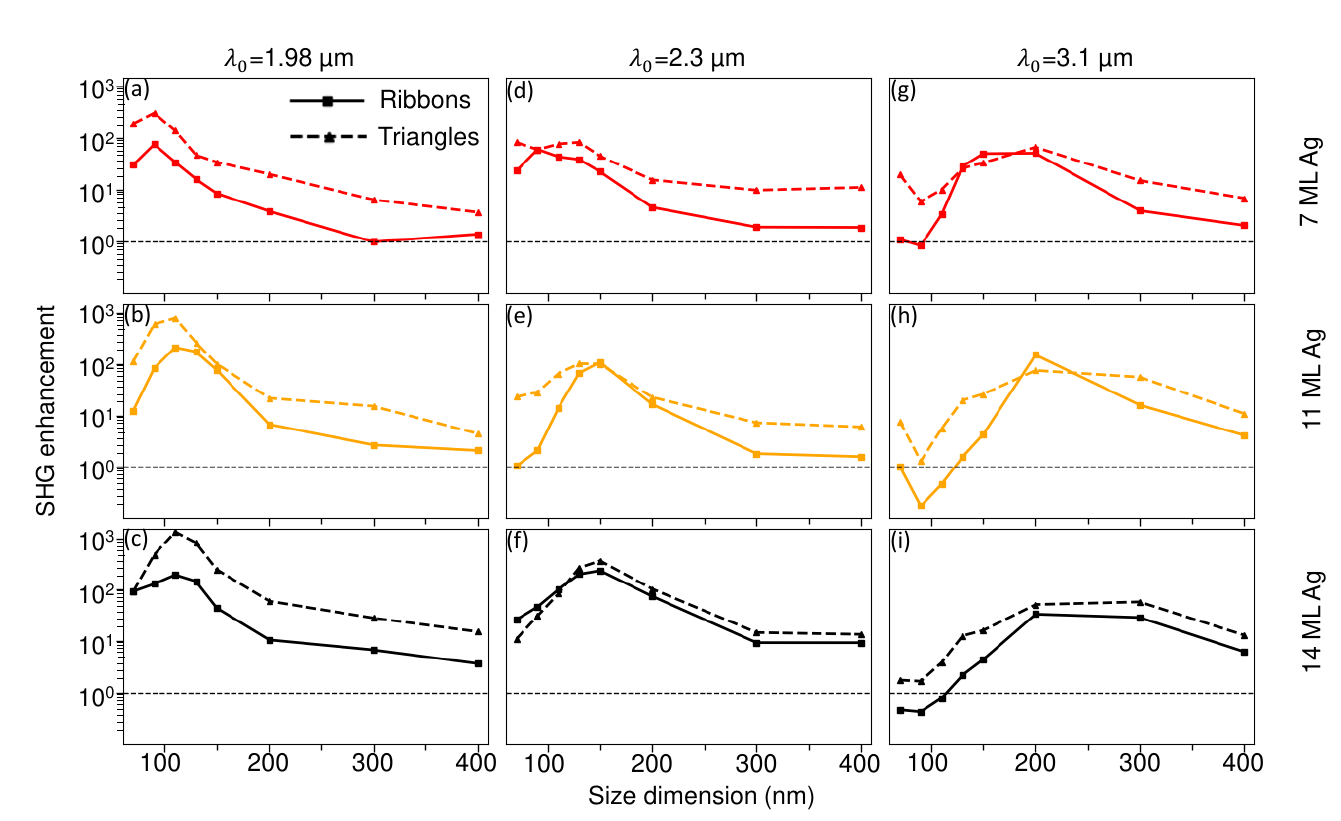}
\caption{\textbf{Comparison of SHG enhancement measured for nanoribbons and triangular nanostructures, normalized by metal filling fraction.} SHG enhancement as a function of lateral size for nanoribbon (solid curves) and nanotriangle (dashed curves) arrays. Each curve is calculated as the ratio between the SHG intensity for the corresponding structured film and that of the unstructured metal film with the same thickness, followed by normalization to the fraction of surface area occupied by metal, so that curves for ribbons are divided by $1/3$ (single periodicity with a period equal to three times the width) and curves for triangles are divided by $\sqrt{3}/36$ (double periodicity with a period equal to three times the side length). Columns correspond to excitation wavelengths $\lambda_0=1.98~\um$ (left), $2.3~\um$ (center), and $3.1~\um$ (right), while rows correspond to different Ag film thicknesses of 7~ML (top), 11~ML (center), and 14~ML (bottom), each of them capped with 4~MLs of Au.}
\label{FigS6}
\end{figure*}

\begin{figure*}[hpbt]
\centering
\includegraphics[width=0.8\linewidth]{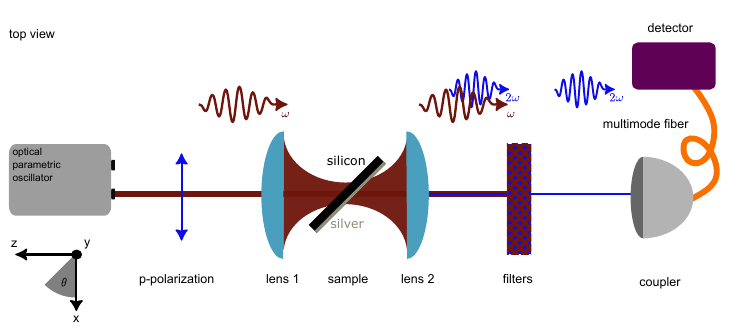}
\caption{\textbf{Sketch of the optical transmission setup.} A Ti:Sapphire oscillator (not shown) pumps an optical parametric oscillator, generating 200-fs p-polarized excitation pulses, which are focused via lens 1 into the $45\,^\circ$-tilted sample. The generated SHG signal is collimated by lens 2, and optical filters spectrally separate it from the excitation light. The isolated SHG signal is then coupled into a multi-mode fiber and directed to the detector.}
\label{FigS7}
\end{figure*}

\begin{figure*}[hpbt]
\centering
\includegraphics[width=1.0\linewidth]{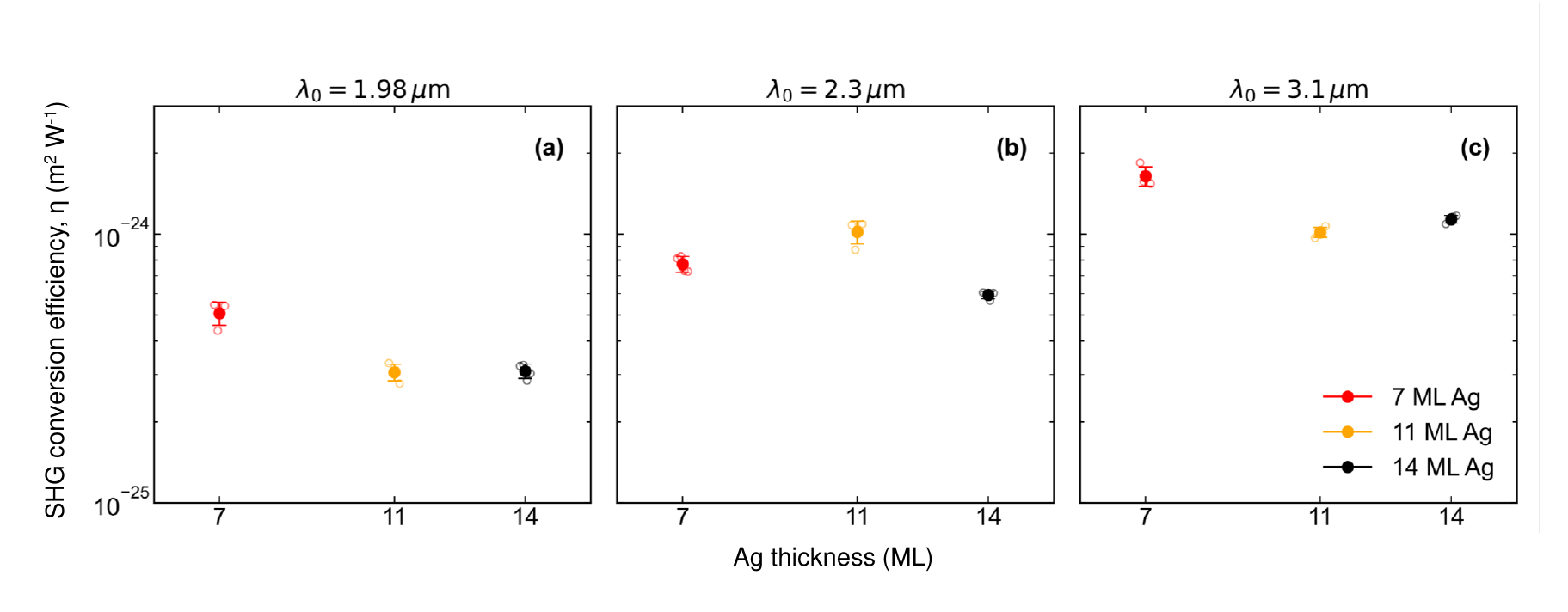}
\caption{\textbf{Thickness-dependent SHG conversion efficiency of unpatterned ultrathin Ag films.} Peak-intensity-normalized SHG conversion efficiency $\eta$ of unpatterned Au-capped ultrathin Ag films as a function of Ag thickness for excitation wavelengths $\lambda_0=1.98~\um$ (a), $2.3~\um$ (b), and $3.1~\um$ (c). For each thickness and excitation wavelength, the SHG signal was measured as a function of the average excitation power. The measured detector power $P_{\mathrm{det}}$ was corrected for the constant background contribution and detector saturation according to $P_{\mathrm{det}}=P_{\mathrm{SHG}}/(1+P_{\mathrm{SHG}}/P_{\mathrm{sat}})+P_{\mathrm{dark}}$, where $P_{\mathrm{SHG}}$ is the corrected average SHG power, $P_{\mathrm{sat}}$ characterizes the detector saturation, and $P_{\mathrm{dark}}$ is the background offset. The corrected SHG power was then fitted to the quadratic dependence $P_{\mathrm{SHG}}=\eta'P_0^2$, yielding the average-power-normalized conversion efficiency $\eta'$. The corresponding average fundamental and SHG powers were converted to peak intensities using the excitation pulse parameters and focal areas, including the Gaussian pulse-shape correction and the reduced SHG beam waist, to obtain $\eta=I_{\mathrm{SHG,peak}}/I_{0,\mathrm{peak}}^2$. Error bars represent the standard deviation extracted from repeated measurements.}
\label{FigS8}
\end{figure*}

\begin{figure*}[hpbt]
\centering
\includegraphics[width=0.9\linewidth]{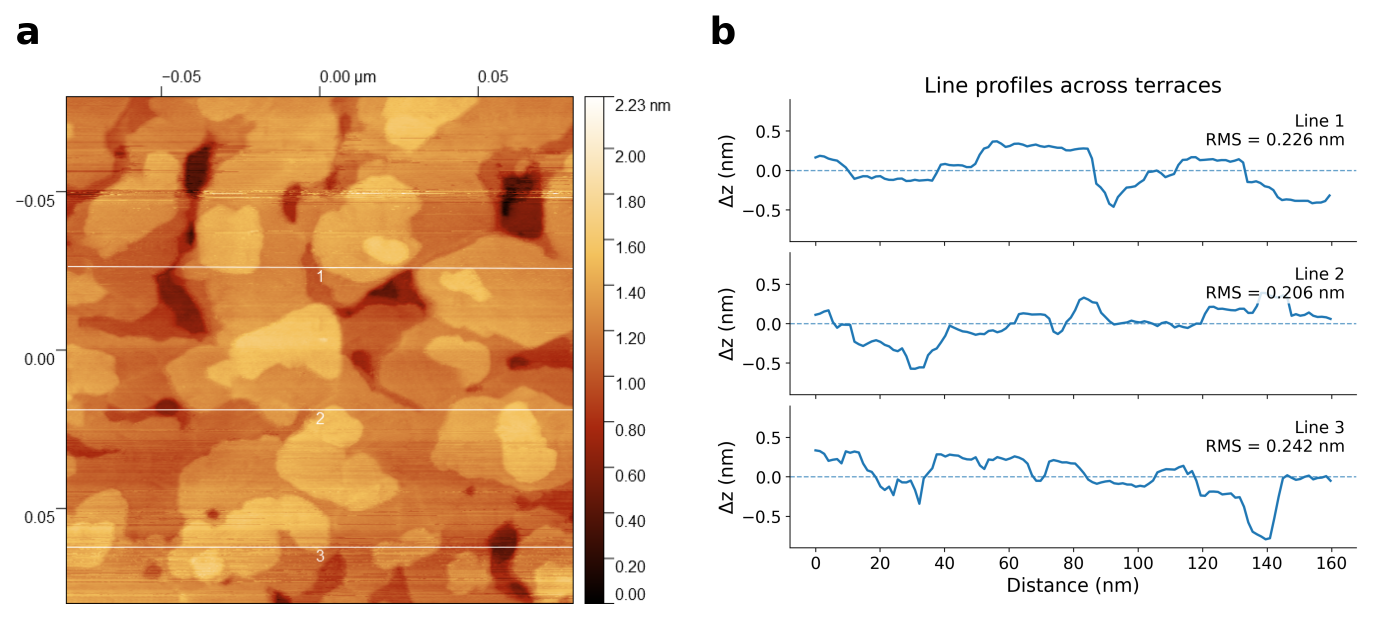}
\caption{\textbf{STM topography and representative height profiles of an Au-capped ultrathin Ag film.} (a)~STM topography of the film surface, revealing a terrace/island-like morphology with local nanoscale height variations on the order of a monolayer. (b)~Corresponding relative-height profiles along the three selected horizontal lines indicated in panel (a), each referenced to its own mean height, together with the extracted one-dimensional RMS roughness values.}
\label{FigS9}
\end{figure*}

\begin{figure*}[hpbt]
\centering
\includegraphics[width=0.95\linewidth]{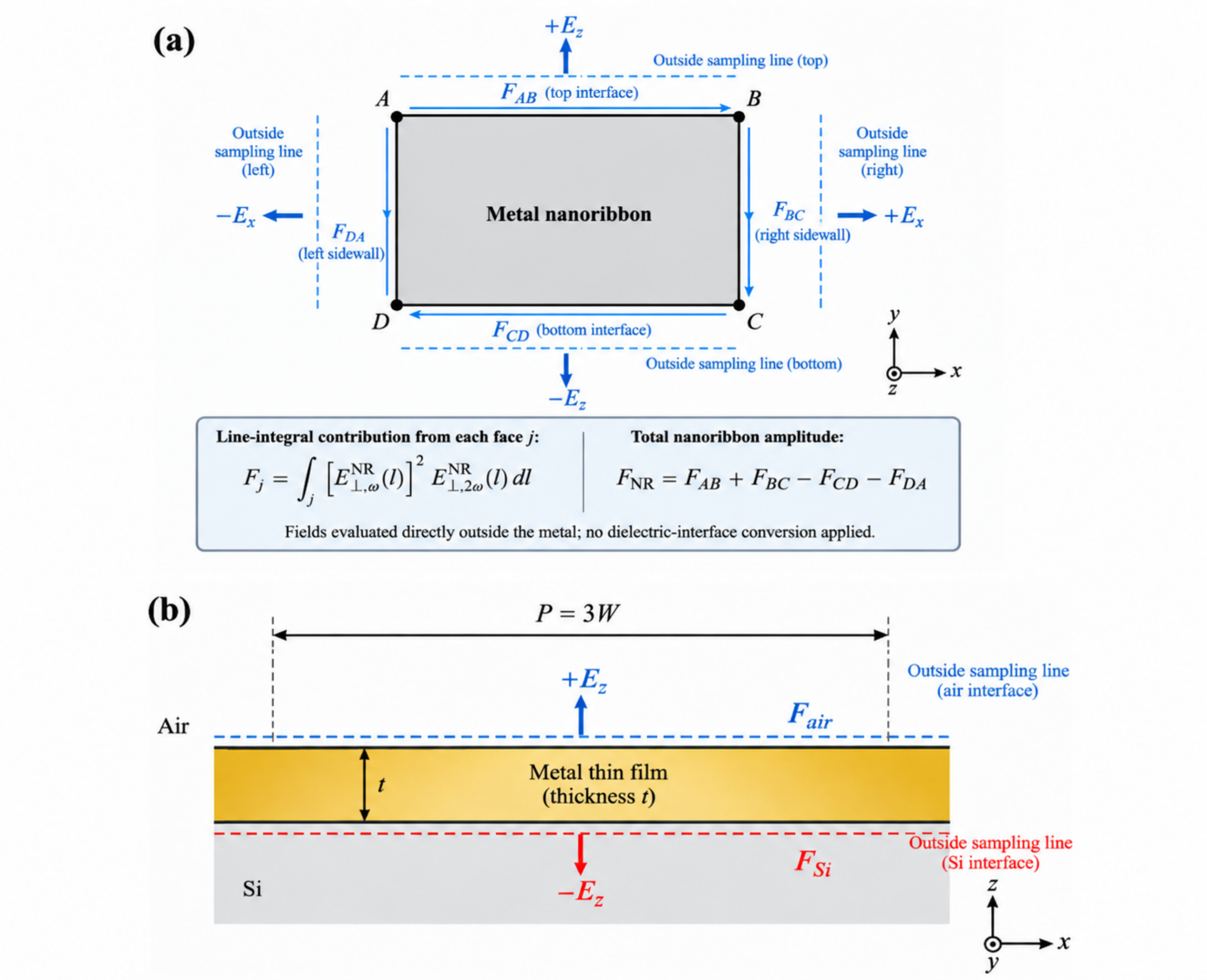}
\caption{\textbf{Schematic of the reciprocity-based surface-integral approach to calculate the SHG enhancement.} The method is based on the SHG surface susceptibility, assuming that the $\chi_{\perp\perp\perp}$ component dominates the second-harmonic response and that this quantity is local and independent of surface orientation. The product of the susceptibility and the square of the normal component of the fundamental field yields the induced second-harmonic surface polarization density, which is likewise directed normal to the surface. By virtue of reciprocity, the far field generated by this polarization density at the detector position is rigorously equal to the surface-normal field induced upon illumination at the second-harmonic frequency $2\omega$ by a plane wave incident from the detector direction. When the ratio of the SHG intensities calculated using this procedure for the patterned and unpatterned films is taken, as in Figure~3 and the Methods section in the main text, the susceptibility cancels, leaving a ratio determined by surface integrals involving the squared normal fundamental field and the corresponding normal second-harmonic field. (a)~Nanoribbon geometry showing the exterior sampling of the surface contour used to evaluate the electric-field component normal to each metal interface at the fundamental frequency $\omega$ and the second-harmonic frequency $2\omega$. The contributions from the individual faces are combined according to the outward-normal convention along the path $A\rightarrow B\rightarrow C\rightarrow D\rightarrow A$. (b)~Corresponding planar thin-film reference, for which the normal electric fields are sampled just outside the air/metal and Si/metal interfaces over a lateral integration window equal to the ribbon-array period $P=3W$. The air- and Si-side contributions are combined using the same outward-normal convention.}
\label{FigS10}
\end{figure*}

\end{document}